\documentclass[12pt,english]{article}
\usepackage{color}
\usepackage{amsmath}
\usepackage{multirow}
\usepackage{graphicx,psfrag,epsf}
\usepackage{enumerate}
\usepackage{natbib}
\newlength\mylen
\usepackage{array} 
\usepackage{listings}
\usepackage{comment}
\usepackage{amssymb}
\usepackage{textcomp}
\usepackage{array,booktabs,arydshln}
\usepackage{tablefootnote}
\usepackage[hyphens]{url}
\usepackage{hyperref}

\newcommand{\bfbeta}{\mbox{\boldmath{$\beta$}}}

\newcommand{\blind}{0}

\usepackage{xr}

\begin{document}

\bibliographystyle{apalike}

\def\spacingset#1{\renewcommand{\baselinestretch}%
{#1}\small\normalsize} \spacingset{1}


\if0\blind
{
  \title{\bf Response Surface Designs for Crossed and Nested Multi-Stratum Structures}
  \author{Luzia A.~Trinca\thanks{
    The first author gratefully acknowledges financial support from FAPESP grant 14/01818-0.}\hspace{.2cm}\\
    Department of Biodiversity and Biostatistics, IBB, Unesp, 
    Brazil\\
    and \\
    Steven G.\ Gilmour\thanks{
    The second author gratefully acknowledges financial support from EPSRC grant EP/T021624/1.}\hspace{.2cm}\\
    Department of Mathematics, King's College London, UK}
  \maketitle
} \fi

\if1\blind
{
  \bigskip
  \bigskip
  \bigskip
  \begin{center}
    {\LARGE\bf Response Surface Designs for Crossed and Nested Multi-Stratum Structures}
\end{center}
  \medskip
} \fi

\bigskip
\begin{abstract}
Response surface designs are usually described as being run under complete randomization of the treatment combinations to the experimental units. In practice, however, it is often necessary or beneficial to run them under some kind of restriction to the randomization, leading to multi-stratum designs. In particular, some factors are often hard to set, so they cannot have their levels reset for each experimental unit. This paper presents a general solution to designing response surface experiments in any multi-stratum structure made up of crossing and/or nesting of unit factors. A stratum-by-stratum approach to constructing designs using compound optimal design criteria is used and illustrated. It is shown that good designs can be found even for large experiments in complex structures.
\end{abstract}

\noindent%
{\it Keywords:} A-optimality; D-optimality; DP-optimality; hard-to-change factor; hard-to-set factor; mixed model; row and column design; strip-block design; strip-plot design.
\vfill

\newpage
\spacingset{2} 

\section{Introduction}
\label{sec:intro}
Many factorial experiments are only possible if the levels of some factors are set less often than others (very hard-to-set, hard-to-set, easy-to-set factors and so on, often called ``hard-to-change'' etc.).
Different levels of hardness-to-set introduce restrictions on the randomization of treatment combinations to runs or experimental units. Each level of hardness-to-set implies a restriction to randomization, which defines a stratum in the analysis from which information arises for parameter estimation. 
Usually, units in a lower stratum are nested within the units in the higher stratum. However, further groupings of the runs to control heterogeneity or to make the experiment feasible lead to more complex configurations, mixing nested and crossed unit factors, potentially in any configuration, each with or without treatment factors applied to them. Common layouts that were given names are blocked, row$\times$column, split-plot, blocked split-plot, and strip-split-plot designs, for example, but many more layouts arise in practice. These appear frequently in industry, agriculture and biotechnology experimental research. A linear mixed model is implied for the analysis of the experimental results. 

For qualitative treatment factors, a balanced structure is often achievable so that constructing an efficient design is not too much of an issue. That is not the case when even a few quantitative factors are present. Many papers on design construction for fitting response surface models in nested multi-stratum structures, especially split-plot and split-split-plot structures, appeared in the last two decades or so, for example, \cite{TrincaGilmour2001, Kowalskietal2006, Goos2006, MylonaGoosJones2014, Samboetal2014, TrincaGilmour2015, TrincaGilmour2017, Borrotietal2017, MylonaGilmouGoos2020,  Borrotietal2023}. 

There has been less emphasis on general crossed and nested structures. In this paper, we consider the construction of efficient designs for quantitative treatment factors under complex unit structures of the design, which involve some crossed unit factors.  To avoid the need for specifying prior information on variance parameters, required for the information matrix under mixed models, and to ensure the designs are efficient for the most challenging cases of larger variance components in the higher strata, we extend the stratum-by-stratum approach proposed previously for dealing with nested structures.

The paper is organized as follows. In Section \ref{sec:unitstruct} the types of design structures tackled in this paper are described and in Section \ref{sec:criteria} the criteria used throughout the rest of the paper are defined. Examples of using the methods in different complex structures, either from real experiments or motivated by and modified from real experiments, are described in Section \ref{sec:examples}. The paper concludes with a discussion in Section \ref{sec:discussion}.

\section{Unit Structures Considered}\label{sec:unitstruct}
Some specific types of crossed-unit structures are well known, especially in the context of unstructured treatments. The Latin square is a famous design; it has two crossed blocking factors, usually called rows and columns, with each combination of a row and a column defining an experimental unit. It has the same number of rows, columns and treatments so that each treatment appears exactly once in each row and exactly once in each column. The generalization of Latin squares to other row$\times$column designs is also well known, especially in agricultural experiments - see, for example, \cite{Mead2012}. Here there can be any number of treatments, rows and columns, so obtaining a design with all treatment effects orthogonally estimated to rows and columns is impossible. Some optimality criterion is needed to find a good design, even for unstructured treatments. For response surface treatment designs, \cite{GilmourTrinca2003} gave an algorithm for arranging optimally, a given treatment design in general row$\times$column design structures. The current paper improves on this by simultaneously optimizing the choice of treatments and their allocation to experimental units in the row$\times$column layout.

Row$\times$column structures are easily generalized to any number of crossed unit factors, though more than two or three is rare in practice. More generally, however, any combination of nesting and crossing is possible. Unit factor $U_2$ is \emph{nested} in unit factor $U_1$ if every level of $U_1$ contains a different set of levels of $U_2$, e.g.\ subplots are nested within whole plots. Unit factors $U_3$ and $U_4$ are \emph{crossed} if every level of $U_3$ appears with every level of $U_4$, e.g.\ rows and columns are crossed. Multiple factors can be nested within a particular factor and crossed with each other. A particular factor can be nested within various other factors which are crossed with each other. 

We will assume that there is a \emph{simple orthogonal unit structure}, i.e.\ there can be any combination of crossing and nesting but, if $U_2$ is nested in $U_1$, each level of $U_1$ contains the same number of $U_2$ units and, if $U_3$ and $U_4$ are crossed, each combination of levels of $U_3$ and $U_4$ appears exactly once. It would be a straightforward extension to our methods to relax these restrictions, but the form of analysis would depend on debatable assumptions, such as the variance component for a unit factor being constant no matter how many of these units appear in each unit of the factor they are nested within or are crossed with. The simple orthogonal block structure ensures that the covariance structure of our linear mixed model is correct, without requiring strong model assumptions.

It is often useful to use the \emph{Wilkinson-Rogers notation}, which denotes nesting and crossing by $/$ and $*$ respectively, e.g.\ $U_1/U_2$ and $U_3*U_4$. We also find it useful to illustrate the unit structure using a \emph{Hasse diagram}. This is a graph having each unit factor represented by a node. Nested factors appear with the nesting factor above the nested factor and their nodes are joined by an edge. Crossed factors occur at the same level in the diagram. They are not joined by an edge but are both joined to a lower factor representing the combinations of their levels and a higher factor representing the higher level units which are divided into rows and columns. We always put a node at the top to represent the entire experiment and one at the bottom to represent the observational units. This will become clearer when we see examples in Section \ref{sec:examples}.

\section{Criteria Considered}\label{sec:criteria}
We assume the model for the response variable measured on the $n$ observational units of the experiment is the linear mixed model \begin{equation}\mathbf{Y}=\mathbf{T}\boldsymbol{\mu}+\mathbf{Z}\mathbf{b}+\boldsymbol{\epsilon},\label{eq:treatmmodel}\end{equation}
where $\mathbf{Y}$ is the $n\times1$ random response vector, $\mathbf{T}$ is the $n\times t$ full treatment indicator matrix $(t<n)$, $\boldsymbol{\mu}$ is the $t\times1$ vector of treatment means, $\mathbf{b}$ is the $u\times 1$ vector of random effects associated with stratum units of the design,  $\mathbf{Z}$ is a $n\times u$ indicator matrix of the units in all strata and $\boldsymbol{\epsilon}$ is the random error vector associated with observational units. Note that a \emph{treatment} here is any combination of levels of the treatment factors which is used in the experiment and $u$ is the total number of random effects incorporated in the $\mathbf{b}$ vector. All random terms are assumed normally distributed with zero means, in particular, $\boldsymbol{\epsilon}\sim N_n(\mathbf{0}, \sigma^2\mathbf{I})$. Furthermore, random terms defined in different strata are assumed to be uncorrelated. 

The $\mathbf{Z}$ matrix and the vector $\mathbf{b}$ are determined by the nesting and crossing structure of the design. For example, for a simple row$\times$column structure with $n_1=r$ rows and $n_2=c$ columns resulting in $n=rc$ units,  $\mathbf{Z}=(\mathbf{Z}_r||\mathbf{Z}_c)$ is an $n\times(r+c)$ matrix, where $\mathbf{Z}_r=\mathbf{1}_c\otimes\mathbf{I}_r$  is an $n\times r$ matrix indicator of rows and $\mathbf{Z}_c=\mathbf{I}_c\otimes\mathbf{1}_r$ is an $n\times c$ matrix indicator of columns. In this case, $\mathbf{b}=(\mathbf{b}_r^\prime,~\mathbf{b}_c^\prime)^\prime$ such that $\mathbf{b}_r\sim N_r(\mathbf{0}, \sigma_r^2\mathbf{I})$, 
$\mathbf{b}_c\sim N_c(\mathbf{0}, \sigma_c^2\mathbf{I})$ and $\mathbf{b}_r$ and $\mathbf{b}_c$ are uncorrelated. For a balanced three-stratum nested structure, with $n_1$ units in the first stratum, each of size $n_2$ units in the second stratum, each of size $n_3$ units in the last stratum, $\mathbf{Z}=(\mathbf{Z}_1||\mathbf{Z}_2)$ such that $\mathbf{Z}_1=\mathbf{1}_{n_2n_3}\otimes\mathbf{I}_{n_1}$ is an $(n_1n_2n_3)\times n_1$ matrix indicator of units in stratum 1 and $\mathbf{Z}_2=\mathbf{1}_{n_3}\otimes\mathbf{I}_{n_2}\otimes\mathbf{I}_{n_1}$ is an $(n_1n_2n_3)\times n_2$ matrix indicator of units in stratum 2. In this case, $\mathbf{b}=(\mathbf{b}_1^\prime,~\mathbf{b}_2^\prime)^\prime$ such that $\mathbf{b}_1\sim N_{n_1}(\mathbf{0};~\sigma_1^2\mathbf{I})$ and $\mathbf{b}_2\sim N_{n_1n_2}(\mathbf{0};~\sigma_2^2\mathbf{I})$. 

For quantitative factors, the fixed part of the equation (\ref{eq:treatmmodel}) is usually approximated by a polynomial model so that the marginal mean vector is 
\begin{equation}
    E(\mathbf{Y})=\mathbf{X}\boldsymbol{\beta}, \label{eq:EY}
\end{equation}
where $\boldsymbol{\beta}$ is a $p\times 1$ vector of fixed coefficients ($p\leq t\leq n$) and $\mathbf{X}$ ($n\times p$) contains columns specifying the effects of interest (e.g.\ linear, quadratic and interaction terms) and the intercept. For known variance components, an estimate of $\bfbeta$ can be obtained by generalized least squares (GLS), namely
\begin{equation}
\widehat{\bfbeta}=(\mathbf{X}^\prime\mathbf{V}^{-1}\mathbf{X})^{-1}\mathbf{X}^\prime\mathbf{V}^{-1}\mathbf{Y}, \label{eq:EstimatorBeta}
\end{equation}
with variance given by
\begin{equation}
\mathbf{V}(\widehat{\bfbeta})=\sigma^2(\mathbf{X}^\prime\mathbf{V}^{-1}\mathbf{X})^{-1}, \label{eq:VarBeta}
\end{equation}
where 
$\mathbf{V}=\mathbf{Z}\boldsymbol{\Psi}^\star\mathbf{Z}^\prime+\mathbf{I}_n$, for $\boldsymbol{\Psi}^\star$ defined as the appropriate block diagonal matrix, with $S$ blocking factors for $S+1$ strata including the lowest error stratum. The $s^{th}$ block matrix is given by 
\[
{\boldsymbol{\Psi}}_s=\frac{\sigma_s^2}{\sigma^2}\mathbf{I}=\eta_s\mathbf{I}
\]
for $s \in \{1,2,\cdots,S\}$.  

The problem of designing based on the mixed model formulation is addressed in the literature, sometimes for the estimation of only fixed effects, and sometimes for the estimation of both fixed effects and variance components. In the first case, the design optimizes a function of the matrix $(\mathbf{X}^\prime\mathbf{V}^{-1}\mathbf{X})$, the so-called fixed-effects optimality. For nested unit structures involving only two strata, e.g.\ blocked or split-plot designs, there are several published papers focusing on the fixed-effects $D$-optimality criterion, which considers local optimization  (point prior information for variance ratios) or a Bayesian framework (incorporating prior probability distributions for the unknown parameters). See, for example, \cite{Goos2002, GoosVandebroek2003, Goos2006, JonesGoos2007, JonesGoos2009, JonesGoos2012, Cuervoetal2017, Caoetal2017}. Mostly, these works deal with relatively small experiments, only two or three nested strata and a few factors to be applied to units in each stratum. There is at least one commercial software package, JMP \citep{JMP}, that, in principle, can construct designs, based on the fixed-effects $D$, $A$ or $I$ criteria, for any unit structure, using the local approach and assuming by default that every variance component is equal to $\sigma^2$. We will call a design which optimizes such a criterion the mixed-model fixed-effects $D$-optimal design.

Other authors considered optimum designs for estimating both types of parameters, optimizing some function of the information matrix for fixed effects and the information matrix for the variance parameters, the last derived from the residual maximum likelihood (REML) estimation method \citep{MylonaGoosJones2014, MylonaGilmouGoos2020}. Again, only two nested strata were considered.

The construction of larger designs for any number of nested strata, using a stratum-by-stratum approach, was proposed in \cite{TrincaGilmour2001, TrincaGilmour2015, TrincaGilmour2017}, with the 2017 paper introducing criteria which optimize for inference, i.e.\ designs that guarantee pure error degrees of freedom in each stratum while maximizing information for the fixed-effects estimation. As noted in \cite{TrincaGilmour2017}, the advantages of the stratum-by-stratum approach are that it is general for any nested multi-stratum structure and does not require prior estimates of variance components. 

In this paper, we show how to implement the approach for complex structures involving nested and crossed design factors. Depending on the structure at each stratum, we should optimize either a completely randomized design, a randomized blocked design, a randomized row$\times$column design, or extensions of the latter for more than two blocking factors. Blocking effects are treated as fixed at the design phase, to ensure designs are efficient in the most challenging case of having large variance components in the higher strata. Thus, we propose compound criteria, as in \cite{GilmourTrinca2012}, to tailor the design according to the experiment's objectives. 
No matter the structure that should be optimized in each phase, the criteria can be written in the general form once we allow all quantities to vary according to the model and structure to be considered in the particular stratum we are designing, that is, we maximize
\begin{equation}
{\frac{|\mathbf{X}_s^\prime\mathbf{Q}_s\mathbf{X}_s|^\frac{\kappa_{D}+\kappa_{DP}}{p_s-1}
(m_s-d_s)^{\kappa_{DF}}} {(F_{p_s-1,d_s;1-\alpha_{DP}})^{\kappa_{DP}}(F_{1,d_s;1-\alpha_{LP}})^{\kappa_{LP}}
[tr\{\mathbf{W}_s(\mathbf{X}_s^\prime\mathbf{Q}_s\mathbf{X}_s)^{-1}\}]^{\kappa_{L}+\kappa_{LP}}}},
\label{eq:CP}
\end{equation}
where the subscript $s$ indicates that the term is stratum specific, $\mathbf{X}_s$ is the approximating model matrix excluding the intercept, $p_s$ is the number of fixed effects parameters, $m_s$ and $d_s$ are, respectively, the number of units and the number of pure error degrees of freedom, $\mathbf{W}_s$ is a diagonal matrix of weights for the weighted-$A$ criterion, $F_{p_s-1,d_s;1-\alpha_{DP}}$ and $F_{1,d_s;1-\alpha_{LP}}$ are the quantiles of the $F$ distributions associated with $(DP)_S$ and $LP$ efficiencies, respectively. The $\kappa$'s are weights associated with the property we desire, i.e. $\kappa_{DP}$ is the weight for criterion $(DP)_S$, $\kappa_{DF}$ is the weight associated with treatment degrees of freedom efficiency and so on. The sum of all weights should add to one, thus $\kappa_{DP}=1$ and all others set to zero results in the $(DP)_S$ criterion that maximizes the determinant and pure error degrees of freedom. 

In each stratum in which there are treatments to be applied to its units, we optimize $\mathfrak{X}_s$, the matrix of level combinations of factors for that stratum, using the point exchange algorithm to maximize the function in (\ref{eq:CP}). For $s>1$, the process maintains fixed the level combinations of factors' $\mathfrak{X}_{j}$s where $j$ ($1\leq j\leq s-1$) belongs to the set of stratum labels that have factors applied to them. For calculating pure error degrees of freedom, $d_s$, we use the model for full treatment effects that would be estimated in stratum $s$, that is, the treatment indicator matrix for factors in stratum $s$ plus all possible interaction terms between these factors and factors applied to previous strata. Let $\mathbf{T}_s$ be the matrix whose columns are indicators for these treatments.

We consider the intercept and blocking fixed effects as nuisance parameters, thus we optimize for $D_S$-, $A_S$-, $(DP)_S$-efficiencies or some compromise between them. Then, the matrix $\mathbf{Q}_s$ which appears in (\ref{eq:CP}) is 
\begin{equation}
\mathbf{Q}_s=\mathbf{I} -\frac{1}{m_s}\mathbf{1}\mathbf{1}^\prime, \label{eq:Qcrd}
\end{equation}
for a completely randomized design, 
\begin{equation}\mathbf{Q}_s=\mathbf{I}-\mathbf{Z}_{b_s}(\mathbf{Z}_{b_s}^\prime\mathbf{Z}_{b_s})^{-1}\mathbf{Z}_{b_s}^\prime, \label{eq:Qrbd}\end{equation} 
for a blocked design where $\mathbf{Z}_{b_s}$ is the $m_s\times b_s$ matrix, whose columns are indicators for the ${b_s}$ blocks, and 
\begin{equation}\mathbf{Q}_s=\mathbf{I}-\mathbf{Z}_{rc}(\mathbf{Z}_{rc}^{\star\prime}\mathbf{Z}_{rc}^\star)^{-1}\mathbf{Z}_{rc}^\prime, \label{eq:Qrc}\end{equation}
for a crossed unit structure such that $\mathbf{Z}_{rc}=(\mathbf{Z}_{r_s}||\mathbf{Z}_{c_s})$ is the $m_s\times (r_s+c_s)$ matrix, with $\mathbf{Z}_{r_s}$ indicators for rows and $\mathbf{Z}_{c_s}$  indicators for columns. Furthermore, $\mathbf{Z}_{rc}^\star$ is $\mathbf{Z}_{rc}$ augmented to take account of the constraints necessary for a unique solution.

\section{Applications} \label{sec:examples}
In this section, we show efficient designs for fixed effects estimation that allow degrees of freedom for pure error, one for a standard row$\times$column design, one for a split-row$\times$column design, one for a strip-split-plot design and one for a split-row$\times$split-column design. We got the motivation from published designs, but since inference is not always possible with tight experiments, we increased their sizes when necessary to allow for pure error degrees of freedom. For comparisons, we obtained designs by methods that do not include degrees of freedom for pure error in the criteria using the modified stratum-by-stratum approach and, when possible, the mixed-model fixed-effects $D$ optimum designs, obtained from JMP \citep{JMP}. Designs for JMP use equal variance components for all strata and designs which optimize the criterion used by JMP are denoted $D^\star$ designs. In all examples, we constructed designs, using our algorithm, based on the $D_S$, $(DP)_S$ and a composite criterion ($CP$) using weights $\boldsymbol{\kappa}=(\kappa_D,~\kappa_{DP},~\kappa_{L},~\kappa_{LP},~\kappa_{DF})=(0,~1/3,~1/3,~0,~1/3)$, balancing among inference ($DP_S$), point estimation ($A_S$) and lack-of-fit degrees of freedom efficiencies. Designs constructed by the stratum-by-stratum approach, optimizing the $D_S$, $(DP)_S$ and $CP_{\boldsymbol{\kappa}}$ criteria, will be referred to as MSS$_{D_S}$, MSS$_{(DP)_S}$ and MSS$_{CP_{\boldsymbol{\kappa}}}$ optimum designs, respectively, where MSS stands for modified stratum-by-stratum.

\subsection{Example 1: pastry dough experiment in a row\texorpdfstring{$\times$}{}column structure}
\begin{figure}
    \centering
    \includegraphics[width=12cm, height=10cm]{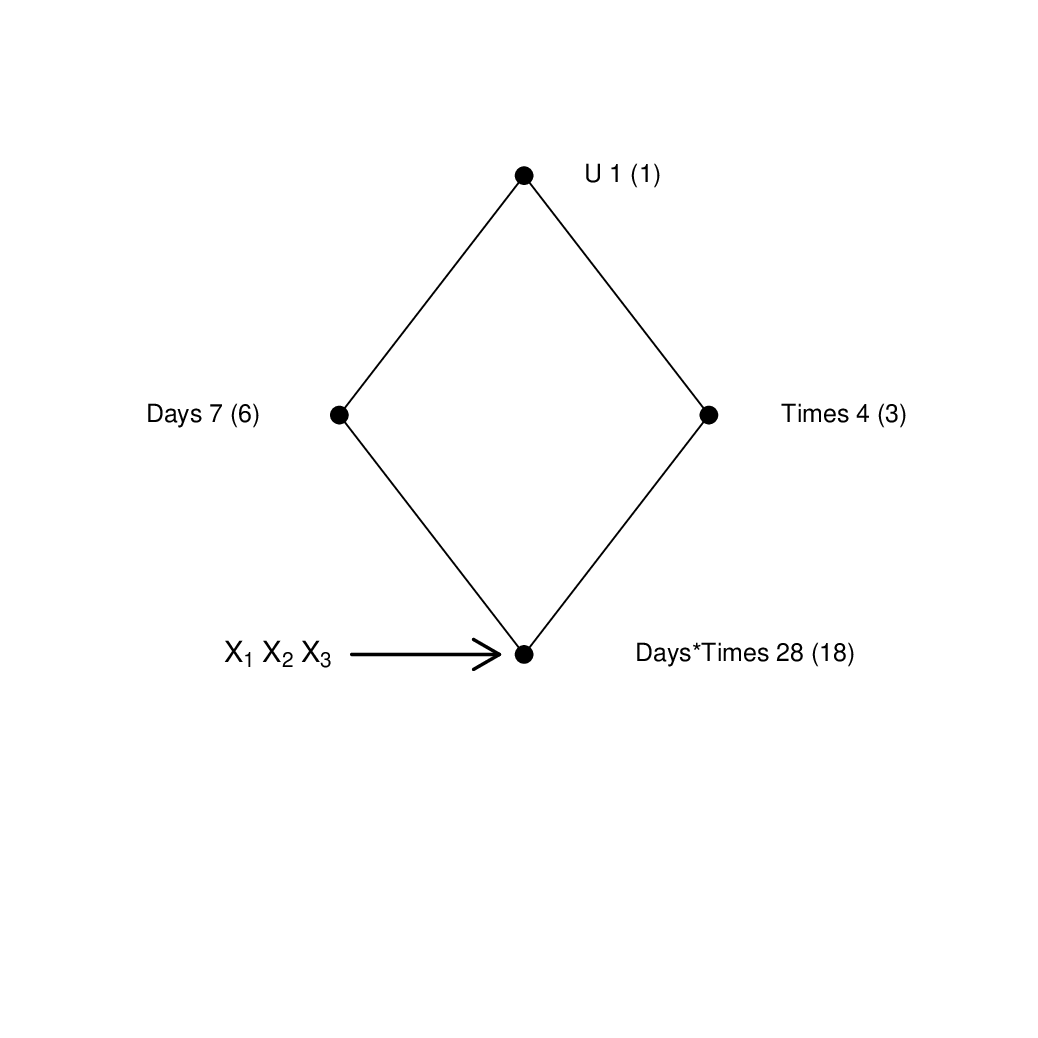}\vspace{-2cm}
    \caption{Hasse diagram of the unit structure of the row$\times$column design for Example 1.}
    \label{fig:hasseEx1}
\end{figure}

In this example, we illustrate our method for constructing row$\times$column designs. These design types occur very often in agriculture where the units are arranged in two-dimensional space and there is the need to control heterogeneity in both directions. However, there are many other areas in which such design types could be efficiently used. \cite{GilmourTrinca2003} discussed their usefulness in the food process industry where, often, experimental units are run one at a time to use the same piece of equipment usually with different settings. The experiment may involve several days and day-to-day variation may be important. Even variation between runs performed on the same day, one at a time, may be important for some processes like, for example, baking. The experiment illustrated in \cite{GilmourTrinca2003} involved three 3-level factors whose effects on several properties of pastry dough were to be studied. Seven days with four runs each were used to experiment. Thus, to control for days and times of day heterogeneity, we designed three 4$\times$7 row$\times$column designs, a $D_S$ optimum design, a $(DP)_S$ optimum design and a compromise design. Figure \ref{fig:hasseEx1} gives the Hasse diagram of the unit structure of the design, showing there are three strata of units, with treatment factors to be applied to the units in the third stratum defined as the Days*Times stratum. Our algorithm to construct these row$\times$column designs is the same as \cite{GilmourTrinca2012} except that the $\mathbf{Q}_s$ matrix used in equation (\ref{eq:CP}) takes account of the adjustment for row and column effects, as defined in equation (\ref{eq:Qrc}).

The designs obtained, using at least 1,000 initial designs, are shown in Table \ref{tab:desEx1}. Table \ref{tab:dfEx1} gives the skeleton analysis of variance, including the strata, sources of variation and degrees of freedom, for each of these designs. It shows the expected pattern of degrees of freedom, with $D_S$ prioritizing treatments, ending up with no PE degrees of freedom, while $(DP)_S$ allows nine PE degrees of freedom but no lack-of-fit checks would be possible. The $D^\star$ design obtained from JMP, using $\eta_R=\eta_C=1$ and 10,000 starting designs, resulted in the same degrees of freedom pattern as our $D_S$ design, as shown in Table \ref{tab:dfEx1}. As for nested design factors \citep{TrincaGilmour2017}, the compound criterion produces useful designs even when the blocking system is more complex. In this example, the design allows seven and two degrees of freedom for pure error and lack of fit, respectively. 

\begin{table}
\caption{Designs for Example 1, a row$\times$column structure Days(7)*Times(4), with three 3-level factors}
\label{tab:desEx1}
\centering
\setlength\tabcolsep{0.15cm}
\renewcommand{\arraystretch}{.7}
\begin{tabular}{c|rrr|rrr|rrr|rrr|rrr|rrr|rrr}
\multicolumn{22}{l}{Design $D^\star$ (JMP)}\\
\multicolumn{22}{c} {~~~~~Days}\\
Times&\multicolumn{3}{c} {1}&\multicolumn{3}{c} {2}&\multicolumn{3}{c} {3}&\multicolumn{3}{c} {4}&\multicolumn{3}{c} {5}&\multicolumn{3}{c} {6}&\multicolumn{3}{c} {7}\\\hline
1	&	0	&	-1	&	-1	&	-1	&	0	&	-1	&	0	&	0	&	0	&	-1	&	-1	&	1	&	1	&	1	&	0	&	1	&	-1	&	1	&	-1	&	1	&	1	\\\hline
2	&	-1	&	1	&	0	&	-1	&	-1	&	1	&	1	&	-1	&	-1	&	1	&	1	&	1	&	-1	&	0	&	-1	&	0	&	1	&	-1	&	1	&	-1	&	1	\\\hline
3	&	1	&	1	&	-1	&	1	&	-1	&	0	&	1	&	1	&	1	&	-1	&	1	&	-1	&	0	&	-1	&	1	&	-1	&	0	&	1	&	-1	&	-1	&	-1	\\\hline
4	&	1	&	0	&	1	&	0	&	1	&	1	&	-1	&	1	&	-1	&	0	&	0	&	0	&	1	&	-1	&	-1	&	-1	&	-1	&	0	&	1	&	1	&	-1	\\\hline

\multicolumn{22}{l}{Design ${D_S}$}\\
\multicolumn{22}{c} {~~~~~Days}\\
Times&\multicolumn{3}{c} {1}&\multicolumn{3}{c} {2}&\multicolumn{3}{c} {3}&\multicolumn{3}{c} {4}&\multicolumn{3}{c} {5}&\multicolumn{3}{c} {6}&\multicolumn{3}{c} {7}\\\hline
1& 1& 1& 1&-1& 1& 1& 0&-1&-1& 1& 1&-1& 1&-1& 1& 0& 0& 0&-1& 0& 1\\\hline
2&-1&-1& 1& 0&-1& 1&-1& 1& 0&-1&-1&-1& 1& 0&-1& 1&-1&-1& 1& 1& 1\\\hline
3&-1& 1&-1&-1& 0&-1& 1& 1&-1& 0& 0& 0& 0& 1& 1&-1&-1& 1& 1&-1& 0\\\hline
4& 1&-1&-1& 1& 1& 0& 1& 0& 1&-1& 1& 1&-1&-1& 0&-1&  0& 1&0&1&-1\\\hline
\multicolumn{22}{l}{Design $(DP)_S$}\\
\multicolumn{22}{c} {~~~~~Days}\\
Times&\multicolumn{3}{c} {1}&\multicolumn{3}{c} {2}&\multicolumn{3}{c} {3}&\multicolumn{3}{c} {4}&\multicolumn{3}{c} {5}&\multicolumn{3}{c} {6}&\multicolumn{3}{c} {7}\\\hline
1&-1& 0& 0&-1&-1&-1&-1&-1& 1& 0& 1& 0& 1&-1&-1&-1& 1& 1& 1&-1& 1\\\hline
2& 1& 1& 1& 1&-1& 0&-1& 1&-1& 1& 1& 1&-1& 1&-1& 0& 0& 1&-1&-1&-1\\\hline
3& 1&-1&-1& 0& 0& 1& 0& 1& 0&-1& 0& 0&-1&-1& 1& 1& 1&-1&-1& 1& 1\\\hline
4&-1&-1& 1& 0& 0& 1&-1& 0& 0& 1&-1&-1& 0& 1& 0& 1&-1& 1& 1& 1&-1\\\hline
\multicolumn{22}{l}{Design $CP_{\boldsymbol{\kappa}}$}\\
\multicolumn{22}{c} {~~~~~Days}\\
Times&\multicolumn{3}{c} {1}&\multicolumn{3}{c} {2}&\multicolumn{3}{c} {3}&\multicolumn{3}{c} {4}&\multicolumn{3}{c} {5}&\multicolumn{3}{c} {6}&\multicolumn{3}{c} {7}\\\hline
1&-1& 0&-1& 1& 0& 0&-1&-1& 1& 0& 1& 0& 1& 1& 1&-1& 1& 1& 0&-1&-1\\\hline
2& 0& 1& 0&-1& 1&-1& 1& 1& 1&-1& 0&-1& 1&-1&-1& 1&-1& 1&-1&-1& 1\\\hline
3&-1& 1& 1& 0&-1&-1& 0& 0& 1& 1& 1&-1&-1& 1&-1&-1&-1& 0& 1& 0& 0\\\hline
4&-1&-1& 0& 1& 1& 1& 1&-1&-1& 1&-1& 1& 0& 0& 1& 1& 1&-1&-1& 1&-1\\\hline
\end{tabular}
\end{table}

\begin{table}
\caption{Skeleton ANOVA of designs for Example 1, a row\texorpdfstring{$\times$}column structure Days(7)*Times(4), with three 3-level factors}
\label{tab:dfEx1}
\centering
\setlength\tabcolsep{0.15cm}
\renewcommand{\arraystretch}{1.0}
\begin{tabular}{llrrr}\hline
& &\multicolumn{3}{c}{Designs}\\\cline{3-5}
Stratum & \multicolumn{1}{l}{Source}&\multicolumn{1}{c}{$D^\star$/$D_S$}&\multicolumn{1}{c}{$(DP)_S$}&\multicolumn{1}{c}{$CP_{\boldsymbol{\kappa}}$}\\\hline
\multicolumn{1}{l}{Days} & \multicolumn{1}{l}{Treat: Lack-of-Fit} & 1 & 1 & 1\\
& \multicolumn{1}{l}{PE} & 5 & 5 & 5\\ \cline{2-5}
& \multicolumn{1}{l}{Total} &6&6&6\\ \hline
\multicolumn{1}{l}{Times}&\multicolumn{1}{l}{Treat: Lack-of-Fit} & 1 & 1 & 0\\
& \multicolumn{1}{l}{PE} & 2 & 2 & 3\\ \cline{2-5}
& \multicolumn{1}{l}{Total}&3&3&3\\\hline
Days*Times & \multicolumn{1}{l}{Treat:}&18&9&11\\
&\multicolumn{1}{l}{~~\textit{2$^{nd}$ order}}&\textit{9}&\textit{9}&\textit{9}\\
&\multicolumn{1}{l}{~~\textit{Lack-of-Fit}}&\textit{9}&\textit{0}&\textit{2}\\
&\multicolumn{1}{l}{PE}&{0~~}&{9}&{7}\\ \cline{2-5}
&\multicolumn{1}{l}{Total} & 18& 18& 18\\ \hline
\multicolumn{1}{l}{Total}&&27&27&27\\\hline
\end{tabular}
\end{table}

Note that the $D^*$ design has 21 treatments randomized to the Days*Times stratum, but only has 18 treatment degrees of freedom in that stratum. This is because one treatment effect is confounded with days and one is confounded with times. Because the design has been chosen to optimize the estimation of the second-order polynomial, the missing effects are higher-order terms. In fact, information about these effects appears in the higher strata, just like inter-block information in an incomplete block design. In principle, these could be used to test for lack of fit in each of the Days and Times strata. These tests are independent of, and testing for lack of fit in a different direction from, the more powerful test that can be done in the Days*Times stratum.  

 Efficiencies, shown in Table \ref{tab:effEx1}, were calculated for different values of the variance components for the random row and column effects model, relative to the $D^\star$ design, which shows that obtaining pure error degrees of freedom costs around 7\% and 11\% $D_S$- and $A_S$-efficiency, respectively, when using a compromise design. In this experiment, the sizes of the variance components have little effect on the comparison between designs.
 
\begin{table}
\caption{$D_S$- and $A_S$-efficiencies, relative to the $D^\star$ design, of the row$\times$column designs for Example 1}
\label{tab:effEx1}
\centering
\setlength\tabcolsep{0.15cm}
\renewcommand{\arraystretch}{.8}
\begin{tabular}{cc rrrcrrr}
  \hline
&&\multicolumn{7}{c}{Criterion}\\\cline{3-9}
&&\multicolumn{3}{c}{$D_S$}&&\multicolumn{3}{c}{$A_S$}\\
&&\multicolumn{3}{c}{Designs}&&\multicolumn{3}{c}{Designs}\\\cline{3-5}\cline{7-9}
$\eta_{Days}$&$\eta_{Times}$&$D_S$&$(DP)_S$&	$CP_{\boldsymbol{\kappa}}$&&$D_S$&$(DP)_S$&	$CP_{\boldsymbol{\kappa}}$\\\hline
1& 1 & 99.87 & 84.49 & 93.23 && 100.46 & 78.46 & 90.67 \\ 
10 & 1 & 99.97 & 83.02 & 92.19 && 100.52 & 76.76 & 89.36 \\ 
100 & 1 & 99.98 & 82.83 & 92.06 && 100.53 & 76.54 & 89.18 \\ 
1 & 10 & 99.70 & 83.65 & 93.04 && 100.27 & 77.41 & 90.41 \\ 
10 & 10 & 99.80 & 82.18 & 91.99 && 100.33 & 75.75 & 89.07 \\ 
100 & 10 & 99.81 & 81.99 & 91.86 && 100.34 & 75.53 & 88.89 \\ 
1 & 100 & 99.68 & 83.55 & 93.01 && 100.25 & 77.28 & 90.37 \\ 
10 & 100 & 99.78 & 82.08 & 91.97 && 100.31 & 75.62 & 89.04 \\ 
100 & 100 & 99.79 & 81.89 & 91.83 && 100.32 & 75.40 & 88.86 \\ 
   \hline
\end{tabular}
\end{table}

\subsection{Example 2: protein extraction experiment in a split-row$\times$column structure}
\label{Ex2}

The objective of this experiment, as first described in \cite{TrincaGilmour2001}, was the extraction of protein from a mixture of two sources $A$ and $B$. The factors thought to affect production were the feed position, the feed flow rate, the gas flow rate and the concentrations of $A$ and $B$. The second-order model was thought of as an approximation to the response function. The experimenters had about 20 days to run the experiment but, realized that if the feed position was to be set for each experimental run, as in a completely randomized design, only one run per day would be possible. This characterized the feed position as a hard-to-set (HS) factor. Once its level was set, two runs could be performed per day. \cite{TrincaGilmour2001} proposed the use of 21 days (21 whole plots), each of size two, with one factor applied to stratum 1 and four easy-to-set (ES) factors to stratum 2. Since there was always one run made in the morning and one in the afternoon, it might have been desirable to allow for systematic differences between times of day, e.g.\ in case runs in the morning tend to give lower responses than runs in the afternoon.

\begin{figure}
    \centering
    \includegraphics[width=12cm, height=10cm]{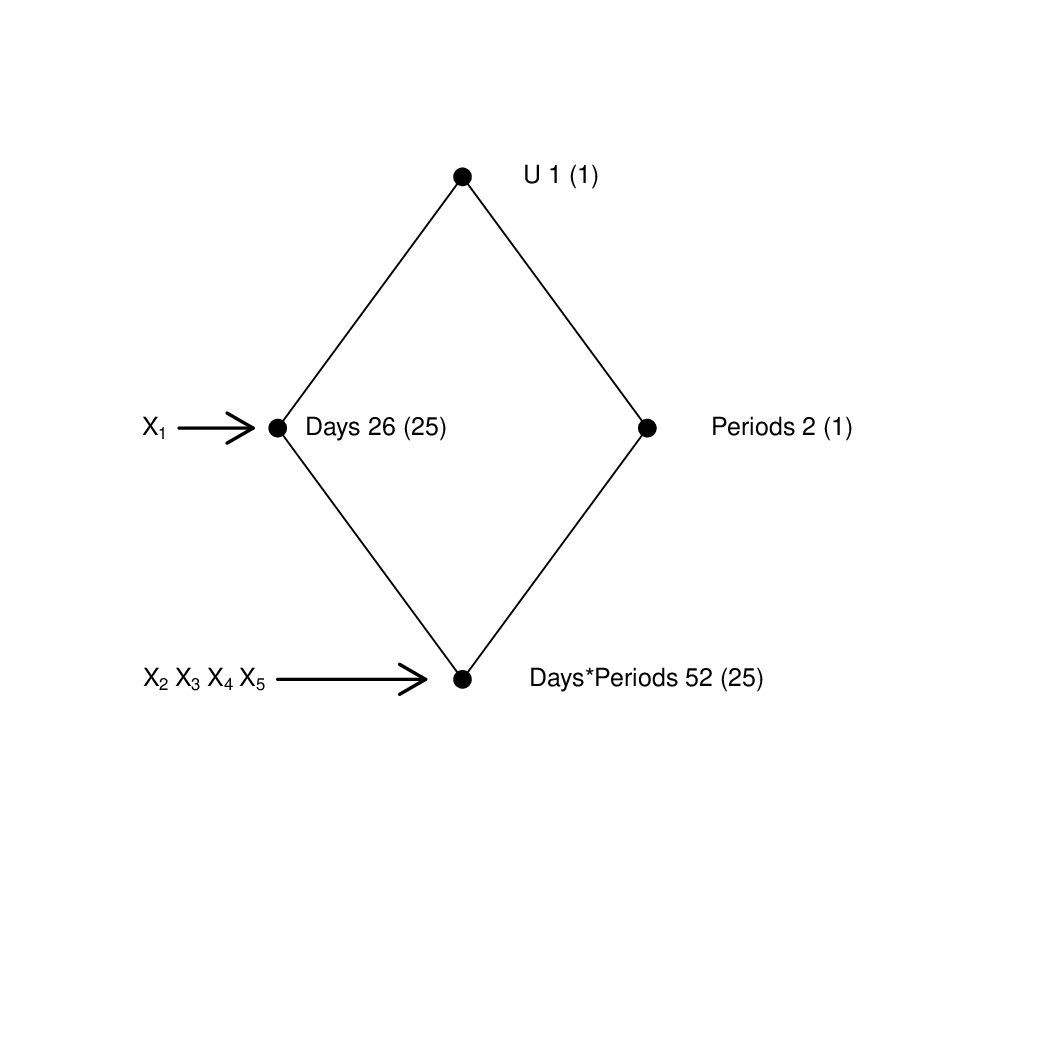}\vspace{-2cm}
    \caption{Hasse diagram of the unit structure of the split-row$\times$column design for Example 2.}
    \label{fig:hasseEx2}
\end{figure}

This would have defined a crossed structure Days*Periods, with the feed position factor applied in the Days stratum, and the other four factors applied in the Days*Periods stratum. Since whole plots are size two and the full second-order model has $p=21$ parameters, the original design was too small to allow inference. Thus we have added five extra days to consider designs which can be used for inference, as in \cite{TrincaGilmour2017}. See Figure \ref{fig:hasseEx2} for the associated Hasse diagram of the unit structure of the design and the factors to be applied to the units of respective strata. Again there are three strata of units with a treatment factor applied to stratum 1, no treatment factor applied to stratum 2 and four treatment factors applied to stratum 3. This type of structure is not one of the common ones given a name. It can be considered a split-plot design with an additional blocking factor crossed with whole plots (days), or a row$\times$column design with an extra treatment factor applied to whole rows. However, fitting a design structure into a restricted class of named designs is unnecessary. The general ideas of crossing and nesting of blocking factors, and careful consideration of which strata treatment factors are applied to, tell us the strategy to optimize the design.

In Tables \ref{tab:designEx2_1} and \ref{tab:designEx2_2} we show four possible designs, the mixed-model fixed-effects $D$-optimum design assuming $\eta_{Days}=\eta_{Periods}=1$ (obtained from JMP) and three designs constructed stratum-by-stratum, using three criterion functions.
For all, we used at least 1,000 initial designs. The steps to build the last three designs, assuming the full three-level factorial points as the candidate set, were:
\begin{enumerate}
        \item Generate a random non-singular design for $X_1$, say $\mathfrak{X}_1^*$, using $m_1=n_1=26$ units. The model includes $p_1=3$ parameters such that $\mathbf{X}_1$ in Equation (\ref{eq:CP}) is $26\times 2$. \label{step:Ex2_1}
    \item Optimize, according to the criterion function in Equation (\ref{eq:CP}), the design $\mathfrak{X}_1^*$ from step \ref{step:Ex2_1} by performing point exchanges. For inference-based criteria, treatment labels are assigned to the rows of $\mathfrak{X}_1^*$ to calculate PE degrees of freedom, where the implicit model is based on the full treatment effects matrix $\mathbf{T}_1$. For inference-based criteria, we use both the treatment-based and the polynomial-based models to optimize the design.
Otherwise, we use just the polynomial-based model matrix. Let $\mathfrak{X}_1^*$ be the optimized
factor levels in stratum 1. The treatments from this matrix will be randomized
to Days or whole plots.
    \item Replicate each row of $\mathfrak{X}_1^*$ twice ($n_2=2$) with each replicate being
assigned to one of the Periods. Let $\mathfrak{X}_1$ be this enlarged matrix with $m_3=n_1n_2=52$ rows.
\item Find a random non-singular initial set of level combinations for $X_2,~\ldots,~X_5$, say $\mathfrak{X}_3$, in a row$\times$column layout considering $n_1=26$ rows (Days) and $n_2=2$ columns (Periods). This is stratum 3 and the polynomial model has $p_3=19$ parameters (the intercept, four linear, four quadratic, and 10 linear-by-linear interactions including those with the factor $X_1$ as set in $\mathfrak{X}_1$) such that $\mathbf{X}_3$ is the $52\times 18$ model matrix for the $p_3-1$ parameters. \label{step:Ex2_2}
\item With $\mathfrak{X}_1$ fixed, optimize, according to the criterion function chosen, the design $\mathfrak{X}_3$ from step \ref{step:Ex2_2}, in the row$\times$column layout, by performing point exchanges involving factors $X_2,~\ldots,~X_5$. For PE degrees of freedom, the model has row and column-blocking effects and the full treatment effects based on $\mathfrak{X}_3$ plus their interactions with full treatment effects based on $\mathfrak{X}_1$, say the indicator matrix $\mathbf{T}_3$. The final design is the optimum design given the initial design.
\item Repeat, from step \ref{step:Ex2_1}, for several initial designs. The best design found among the repetitions is declared to be the optimum.
\end{enumerate}

\settowidth\mylen{27}
\begin{table}
\caption{Designs for Example 2, a split-row$\times$column structure Days(26)*Periods(2), with 1 HS and 4 ES three-level factors}
\label{tab:designEx2_1}
\centering
\setlength\tabcolsep{0.15cm}
\tiny{
\begin{tabular}{ccc c ccc}
 \multicolumn{3}{c}{$D^\star$}&&\multicolumn{3}{c}{MSS$_{D_S}$}\\
&\multicolumn{1}{c}{Morning}&\multicolumn{1}{c}{Afternoon}&&&\multicolumn{1}{c}{Morning}&\multicolumn{1}{c}{Afternoon}\\
$\begin{array}{cr}	Day&X_1\\				
1		&	-1	\\
2		&	-1	\\
3		&	-1	\\
4		&	-1	\\
5		&	-1	\\
6		&	-1	\\
7		&	-1	\\
8		&	-1	\\
9		&	-1	\\
10		&	-1	\\
11		&	-1	\\
12		&	0	\\
13		&	0	\\
14		&	0	\\
15		&	0	\\
16		&	1	\\
17		&	1	\\
18		&	1	\\
19		&	1	\\
20		&	1	\\
21		&	1	\\
22		&	1	\\
23		&	1	\\
24		&	1	\\
25		&	1	\\
26		&	1	\end{array}$ &\hspace{-.3cm}
 $\begin{array}{rrr *{1}{wr{\mylen}}}X_2&X_3&X_4&X_5\\	\hline					
-1 & -1 & -1 & -1 \\ 
  1 & -1 & 1 & -1 \\ 
  -1 & 1 & 1 & -1 \\ 
  1 & 1 & 1 & 1 \\ 
  1 & -1 & -1 & -1 \\ 
  1 & 1 & -1 & -1 \\ 
  0 & -1 & -1 & 1 \\ 
  -1 & 0 & 1 & 1 \\ 
  1 & -1 & 0 & 1 \\ 
  -1 & 0 & 0 & -1 \\ 
  0 & -1 & 1 & 0 \\ 
  0 & 1 & 0 & 0 \\ 
  -1 & -1 & 1 & 1 \\ 
  1 & 1 & -1 & 1 \\ 
  0 & 1 & 1 & -1 \\ 
  1 & 1 & -1 & 1 \\ 
  -1 & 0 & 1 & 0 \\ 
  -1 & 1 & -1 & -1 \\ 
  -1 & -1 & -1 & 1 \\ 
  1 & 0 & 1 & -1 \\ 
  1 & -1 & 0 & -1 \\ 
  1 & 1 & 1 & -0.11 \\ 
  1 & -1 & 1 & 1 \\ 
  1 & -1 & -1 & 0 \\ 
  -1 & -1 & 1 & -1 \\ 
  -1 & 1 & -1 & 1 \\	\end{array}$&\hspace{-.3cm}
	$\begin{array}{rrrr}X_2&X_3&X_4&X_5\\	\hline											
1 & 1 & 1 & -1 \\ 
  -1 & 1 & 1 & 1 \\ 
  1 & -1 & 1 & 1 \\ 
  -1 & -1 & -1 & 1 \\ 
  -1 & 1 & -1 & 1 \\ 
  0 & -1 & 1 & 0 \\ 
  -1 & -1 & 1 & -1 \\ 
  -1 & 1 & -1 & -1 \\ 
  0 & 0 & -1 & -1 \\ 
  1 & 1 & -1 & 0 \\ 
  1 & 0 & -1 & 1 \\ 
  1 & -1 & 1 & -1 \\ 
  1 & -1 & -1 & 0 \\ 
  -1 & 0 & -1 & -1 \\ 
  -1 & 0 & 0 & 0 \\ 
  -1 & 1 & 1 & 1 \\ 
  1 & -1 & -1 & 1 \\ 
  1 & 1 & 1 & 1 \\ 
  1 & 1 & -1 & -1 \\ 
  -1 & -1 & 1 & 1 \\ 
  -1 & 1 & 1 & -1 \\ 
  0 & 0 & -1 & 1 \\ 
  -1 & -1 & -1 & -1 \\ 
  0 & 1 & 0 & 1 \\ 
  1 & 1 & 0 & -1 \\ 
  0 & -1 & 0 & -1 \\ \end{array}$&&\hspace{-.5cm}
$\begin{array}{cr}	Day&X_1\\				
	1		&	-1	\\
	2		&	-1	\\
	3		&	-1	\\
	4		&	-1	\\
	5		&	-1	\\
	6		&	-1	\\
	7		&	-1	\\
	8		&	-1	\\
	9		&	-1	\\
	10		&	0	\\
	11		&	0	\\
	12		&	0	\\
	13		&	0	\\
	14		&	0	\\
	15		&	0	\\
	16		&	0	\\
	17		&	0	\\
	18		&	1	\\
	19		&	1	\\
	20		&	1	\\
	21		&	1	\\
	22		&	1	\\
	23		&	1	\\
	24		&	1	\\
	25		&	1	\\
	26		&	1	\end{array}$&\hspace{-.3cm}
	$\begin{array}{rrrr}X_2&X_3&X_4&X_5\\	\hline							
	-1	&	-1	&	1	&	-1	\\
	1	&	1	&	1	&	1	\\
	1	&	1	&	-1	&	1	\\
	-1	&	-1	&	0	&	1	\\
	1	&	1	&	-1	&	-1	\\
	-1	&	-1	&	-1	&	-1	\\
	-1	&	0	&	-1	&	-1	\\
	-1	&	1	&	1	&	1	\\
	1	&	-1	&	-1	&	0	\\
	0	&	-1	&	1	&	1	\\
	1	&	1	&	1	&	-1	\\
	1	&	-1	&	1	&	-1	\\
	1	&	-1	&	0	&	-1	\\
	1	&	0	&	0	&	-1	\\
	-1	&	1	&	1	&	0	\\
	0	&	1	&	0	&	0	\\
	0	&	0	&	1	&	0	\\
	-1	&	1	&	1	&	-1	\\
	1	&	-1	&	1	&	1	\\
	0	&	-1	&	1	&	-1	\\
	-1	&	1	&	-1	&	-1	\\
	-1	&	1	&	-1	&	1	\\
	-1	&	-1	&	1	&	1	\\
	1	&	0	&	-1	&	1	\\
	1	&	1	&	-1	&	-1	\\
	1	&	-1	&	-1	&	1	\end{array}$&\hspace{-.3cm}	
	$\begin{array}{rrrr}X_2&X_3&X_4&X_5\\		\hline						
	1	&	-1	&	-1	&	1	\\
	-1	&	-1	&	-1	&	1	\\
	-1	&	1	&	1	&	-1	\\
	1	&	-1	&	-1	&	-1	\\
	1	&	-1	&	1	&	1	\\
	1	&	1	&	1	&	-1	\\
	-1	&	-1	&	1	&	1	\\
	1	&	-1	&	1	&	-1	\\
	-1	&	1	&	-1	&	1	\\
	1	&	0	&	0	&	0	\\
	0	&	0	&	0	&	0	\\
	0	&	1	&	-1	&	-1	\\
	-1	&	0	&	-1	&	0	\\
	1	&	1	&	-1	&	1	\\
	0	&	0	&	0	&	-1	\\
	-1	&	0	&	1	&	1	\\
	-1	&	1	&	0	&	-1	\\
	1	&	-1	&	-1	&	-1	\\
	-1	&	-1	&	-1	&	-1	\\
	1	&	1	&	1	&	1	\\
	0	&	1	&	1	&	1	\\
	-1	&	-1	&	1	&	-1	\\
	0	&	1	&	1	&	-1	\\
	1	&	-1	&	1	&	0	\\
	-1	&	-1	&	-1	&	1	\\
	-1	&	1	&	1	&	1	\end{array}$
\\
\end{tabular}}
\end{table}

\begin{table}\caption{Table \ref{tab:designEx2_1} continued}\label{tab:designEx2_2}
\centering
\tiny{
\begin{tabular}{ccc c ccc}
 \multicolumn{3}{c}{MSS$_{(DP)_S}$}&&\multicolumn{3}{c}{MSS$_{CP_{\boldsymbol{\kappa}}}$}\\
&Morning&Afternoon&&&Morning&Afternoon\\
$\begin{array}{cr}	Day&X_1\\				
	1		&	-1	\\
	2		&	-1	\\
	3		&	-1	\\
	4		&	-1	\\
	5		&	-1	\\
	6		&	-1	\\
	7		&	-1	\\
	8		&	-1	\\
	9		&	-1	\\
	10		&	0	\\
	11		&	0	\\
	12		&	0	\\
	13		&	0	\\
	14		&	0	\\
	15		&	0	\\
	16		&	0	\\
	17		&	0	\\
	18		&	1	\\
	19		&	1	\\
	20		&	1	\\
	21		&	1	\\
	22		&	1	\\
	23		&	1	\\
	24		&	1	\\
	25		&	1	\\
	26		&	1	\end{array}$&\hspace{-.3cm}
$\begin{array}{rrrr}X_2&X_3&X_4&X_5\\		\hline						
	1	&	1	&	0	&	-1	\\
	-1	&	0	&	-1	&	-1	\\
	0	&	-1	&	0	&	1	\\
	1	&	1	&	1	&	0	\\
	1	&	1	&	1	&	1	\\
	-1	&	-1	&	-1	&	1	\\
	1	&	-1	&	-1	&	0	\\
	-1	&	0	&	-1	&	-1	\\
	0	&	-1	&	0	&	1	\\
	-1	&	0	&	-1	&	1	\\
	1	&	0	&	0	&	-1	\\
	0	&	1	&	0	&	0	\\
	0	&	1	&	1	&	-1	\\
	-1	&	-1	&	1	&	1	\\
	1	&	0	&	0	&	-1	\\
	1	&	0	&	0	&	-1	\\
	1	&	1	&	-1	&	1	\\
	1	&	1	&	-1	&	-1	\\
	1	&	1	&	-1	&	-1	\\
	-1	&	1	&	1	&	-1	\\
	1	&	1	&	-1	&	-1	\\
	1	&	-1	&	1	&	-1	\\
	1	&	1	&	0	&	1	\\
	-1	&	1	&	-1	&	-1	\\
	1	&	1	&	0	&	1	\\
	0	&	-1	&	-1	&	-1	\end{array}$	
	&\hspace{-.3cm}
	$\begin{array}{rrrr}X_2&X_3&X_4&X_5\\	\hline							
	0	&	-1	&	1	&	0	\\
	-1	&	1	&	0	&	1	\\
	-1	&	1	&	1	&	0	\\
	-1	&	1	&	-1	&	1	\\
	-1	&	1	&	1	&	-1	\\
	1	&	-1	&	1	&	-1	\\
	-1	&	0	&	1	&	1	\\
	-1	&	1	&	0	&	1	\\
	-1	&	1	&	1	&	0	\\
	0	&	1	&	-1	&	-1	\\
	-1	&	-1	&	0	&	-1	\\
	1	&	-1	&	-1	&	1	\\
	1	&	0	&	0	&	0	\\
	0	&	0	&	0	&	-1	\\
	-1	&	-1	&	0	&	-1	\\
	-1	&	-1	&	0	&	-1	\\
	1	&	-1	&	-1	&	-1	\\
	1	&	-1	&	1	&	1	\\
	1	&	-1	&	1	&	1	\\
	-1	&	-1	&	-1	&	1	\\
	1	&	-1	&	1	&	1	\\
	-1	&	1	&	1	&	1	\\
	-1	&	0	&	0	&	0	\\
	-1	&	-1	&	1	&	-1	\\
	-1	&	0	&	0	&	0	\\
	1	&	1	&	1	&	-1	\end{array}$&&\hspace{-.5cm}
	$\begin{array}{cr}	Day&X_1\\				
	1		&	-1	\\
	2		&	-1	\\
	3		&	-1	\\
	4		&	-1	\\
	5		&	-1	\\
	6		&	-1	\\
	7		&	-1	\\
	8		&	-1	\\
	9		&	-1	\\
	10		&	0	\\
	11		&	0	\\
	12		&	0	\\
	13		&	0	\\
	14		&	0	\\
	15		&	0	\\
	16		&	0	\\
	17		&	0	\\
	18		&	1	\\
	19		&	1	\\
	20		&	1	\\
	21		&	1	\\
	22		&	1	\\
	23		&	1	\\
	24		&	1	\\
	25		&	1	\\
	26		&	1	
\end{array}$&\hspace{-.3cm}
$\begin{array}{rrrr}X_2&X_3&X_4&X_5\\		\hline						
	0	&	0	&	0	&	1	\\
	1	&	1	&	1	&	1	\\
	-1	&	1	&	1	&	1	\\
	1	&	1	&	1	&	1	\\
	-1	&	1	&	-1	&	-1	\\
	1	&	-1	&	-1	&	1	\\
	-1	&	1	&	1	&	-1	\\
	1	&	1	&	-1	&	1	\\
	0	&	0	&	0	&	1	\\
		1	&	-1	&	-1	&	-1	\\
	-1	&	-1	&	0	&	1	\\
	-1	&	-1	&	1	&	-1	\\
	0	&	1	&	1	&	0	\\
	-1	&	-1	&	-1	&	-1	\\
	1	&	-1	&	-1	&	-1	\\
	-1	&	-1	&	-1	&	-1	\\
	-1	&	-1	&	0	&	1\\
	0	&	0	&	-1	&	0	\\
	0	&	-1	&	1	&	1	\\
	1	&	1	&	0	&	0	\\
	-1	&	0	&	1	&	1	\\
	-1	&	1	&	-1	&	-1	\\
	0	&	0	&	-1	&	0	\\
	-1	&	1	&	-1	&	1	\\
	1	&	0	&	1	&	-1	\\
	1	&	1	&	-1	&	-1	
	\end{array}$&\hspace{-.3cm}
	$\begin{array}{rrrr}X_2&X_3&X_4&X_5\\		\hline						
	1	&	1	&	0	&	-1	\\
	-1	&	1	&	-1	&	1	\\
	0	&	-1	&	-1	&	-1	\\
	-1	&	1	&	-1	&	1	\\
	1	&	-1	&	1	&	-1	\\
	0	&	0	&	0	&	-1	\\
	1	&	1	&	-1	&	1	\\
	-1	&	-1	&	1	&	1	\\
	1	&	1	&	0	&	-1	\\
1	&	1	&	1	&	-1	\\
	1	&	-1	&	1	&	1	\\
	1	&	0	&	-1	&	0	\\
	1	&	-1	&	0	&	1	\\
	0	&	1	&	0	&	-1	\\
	1	&	1	&	1	&	-1	\\
	0	&	1	&	0	&	-1	\\
	1	&	-1	&	1	&	1\\	
	0	&	1	&	1	&	1	\\
	-1	&	0	&	1	&	-1	\\
	0	&	-1	&	-1	&	1	\\
	0	&	-1	&	-1	&	-1	\\
	1	&	1	&	-1	&	1	\\
	0	&	1	&	1	&	1	\\
	-1	&	-1	&	1	&	0	\\
	-1	&	1	&	-1	&	0	\\
	1	&	-1	&	0	&	0	
		\end{array}$\\
\end{tabular}}
\end{table}

The decomposition of the degrees of freedom for the four designs is shown in Table \ref{tab:dfEx2}, where the usual pattern is observed, with no PE degrees of freedom in both strata when using criteria that do not include this requirement, even in the stratum for Days which includes many units for just one factor, while using $(DP)_S$ or composite criteria allow seven PE degrees of freedom in that stratum. However, all available degrees of freedom in the Days*Periods stratum are allocated to PE for the MSS$_{(DP)_S}$ design. The compromise design allows one degree of freedom for lack of fit instead. The difference between the residual degrees of freedom and the PE degrees of freedom in the Days stratum is not accounted for by lack of fit of the model for factors applied in this stratum but represents inter-days (inter-block) information on treatment effects related to factors applied in the Runs stratum.

\begin{table}
\caption{Skeleton ANOVA of designs for Example 2, a split-row$\times$column structure Days(26)*Periods(2), with 1 HS and 4 ES three-level factors}
\label{tab:dfEx2}
\centering
\setlength\tabcolsep{0.15cm}
\renewcommand{\arraystretch}{.8}
\begin{tabular}{llrrrr}\hline
& &\multicolumn{4}{c}{Designs}\\\cline{3-6}
\multicolumn{1}{l}{Stratum}&\multicolumn{1}{l}{Source}&\multicolumn{1}{c}{$D^\star$}&\multicolumn{1}{c}{MSS$_{D_S}$}&\multicolumn{1}{c}{MSS$_{(DP)_S}$}&\multicolumn{1}{c}{MSS$_{CP_{\boldsymbol{\kappa}}}$}\\\hline
Days&\multicolumn{1}{l}{Treat ($\mathbf{T}_1$): $X_1$, $X_1^2$}&2&2&2&2\\
&Treat ($\mathbf{T}_3$): Lack-of-Fit&23&23&16&16\\
&\multicolumn{1}{l}{PE}&0&0&7&7\\\cline{2-6}
& \multicolumn{1}{l}{Total}& 25&25&25&25\\\hline
Periods& &1&1&1&1\\\hline
Runs&\multicolumn{1}{l}{Treat ($\mathbf{T}_3$):}&25&25&18&19\\
&\multicolumn{1}{l}{~~\textit{2$^{nd}$ order}}&\textit{18}&\textit{18}&\textit{18}&\textit{18}\\
&\multicolumn{1}{l}{~~\textit{Lack-of-Fit}}&\textit{7}&\textit{7}&\textit{0}&\textit{1}\\
&\multicolumn{1}{l}{PE}&0&0&7&6\\ \cline{2-6}
&\multicolumn{1}{l}{Total} & 25 & 25 & 25 & 25\\ \hline
Total& &51&51&51&51\\\hline
\end{tabular}
\end{table}

The $D_S$- and $A_S$-efficiencies, relative to the mixed-model fixed-effects $D$-optimum design, are shown in Table \ref{tab:effEx2}. It shows that in this case, the mixed-model fixed-effects $D$-optimum design depends on the variance component ratios since the MSS$_{D_s}$ design has efficiencies greater than 100\% as the day-to-day variability increases. The gain in efficiencies of the stratum-by-stratum constructed designs is larger for the $A_S$ criterion. 

\begin{table}
\caption{$D_S$- and $A_S$-efficiencies, relative to the $D^\star$ design, of the split-row$\times$column designs for Example 2}\label{tab:effEx2}
\centering
\setlength\tabcolsep{0.15cm}
\renewcommand{\arraystretch}{.8}
\begin{tabular}{cc rrrccrrr}\hline
&&\multicolumn{8}{c}{Criterion}\\\cline{3-10}
&&\multicolumn{3}{c}{$D_S$}&&&\multicolumn{3}{c}{$A_S$}\\
&&\multicolumn{3}{c}{Designs}&&&\multicolumn{3}{c}{Designs}\\\cline{3-5}\cline{8-10}
$\eta_{Day}$&$\eta_{Period}$&\multicolumn{1}{c}{MSS$_{D_S}$}&\multicolumn{1}{c}{MSS$_{(DP)_S}$}&\multicolumn{1}{c}{MSS$_{CP_{\boldsymbol{\kappa}}}$}&&&\multicolumn{1}{c}{MSS$_{D_S}$}&\multicolumn{1}{c}{MSS$_{(DP)_S}$}&\multicolumn{1}{c}{MSS$_{CP_{\boldsymbol{\kappa}}}$}\\\hline
  1 & 1 & 97.67~~& 81.67~~~~& 86.56~~&&& 96.62~~& 76.92~~~~& 84.21~~ \\ 
 10 & 1 & 100.97~~& 79.02~~~~& 86.30~~&&& 110.01~~& 89.59~~~~& 98.61~~  \\ 
100 & 1 & 101.81~~& 78.26~~~~& 86.27~~&&& 117.13~~& 111.76~~~~& 114.46~~  \\ 
  1 & 10 & 97.66~~& 81.56~~~~& 86.51~~&&& 96.61~~& 76.81~~~~& 84.16~~ \\ 
 10 & 10 & 100.95~~& 78.89~~~~& 86.24~~&&& 110.00~~& 89.49~~~~& 98.55~~ \\ 
100 & 10 & 101.80~~& 78.13~~~~& 86.20~~&&& 117.13~~& 111.72~~~~& 114.45~~ \\ 
  1 & 100 & 97.66~~& 81.55~~~~& 86.50~~&&& 96.61~~& 76.80~~~~& 84.15~~  \\ 
 10 & 100 & 100.95~~& 78.88~~~~& 86.23~~&&& 110.00~~& 89.47~~~~& 98.55~~ \\ 
100 & 100 & 101.80~~& 78.11~~~~& 86.20~~&&& 117.13~~& 111.72~~~~& 114.45~~  \\ \hline
\end{tabular}
\end{table}

Furthermore, these designs seem less dependent on the variance component ratios, in terms of $D_S$-efficiencies. The requirement of PE degrees of freedom in the MSS$_{(DP)_S}$ and in the compromise designs causes a reduction of about 20\% and 14\%, respectively, in terms of $D_S$. There are much larger gains in terms of $A_S$-efficiency as day-to-day variation increases. 

\subsection{Example 3: strip-split-plot design}

\begin{figure}
    \centering
    \includegraphics[width=12cm, height=10cm]{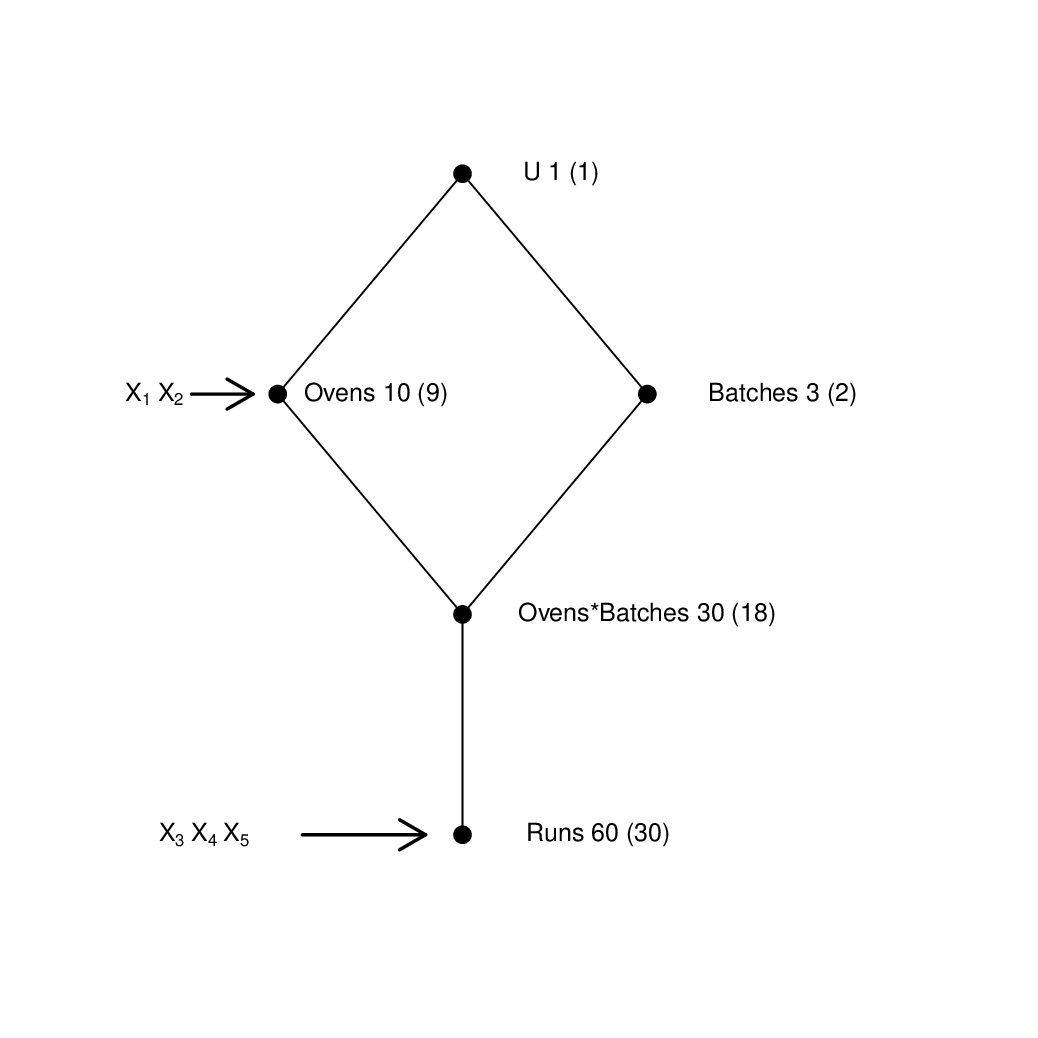}\vspace{-1cm}
    \caption{Hasse diagram of the unit structure of the strip-split-plot design for Example 3.}
    \label{fig:hasseEx3}
\end{figure}

This example is motivated by \cite{JensenKowalski2012} who described an experiment for the production of some manufactured components. The manufacturing process involved the components being treated in the oven under level combinations of two factors, time ($X_1$) and temperature ($X_2$). For each setting of these two factors six component units would go together inside the oven and ten oven runs would be performed. Each component unit could be further subjected to levels of another factor under investigation, the orientation inside the oven ($X_3$). Due to material limitations, the components inside the oven would belong to three different batches, so blocking was required. Thus, there were 10 whole rows each of size six, these whole rows being crossed with three columns each of size 20. In total there were 60 units and whole-rows and columns have a crossed layout, i.e.\ (Ovens*Batches)/Runs. The original experiment, involving two three-level HS factors (temperature and time settings for the oven) and one two-level ES factor, was quite a large experiment to investigate a few effects. To challenge our procedure we used the same unit layout but applied three three-level factors to the units within whole-plot$\times$batch combinations. Figure \ref{fig:hasseEx3} shows the Hasse diagram for the unit structure of the design and the factors to be applied to the units of each stratum.
There are four strata but only stratum 1 and stratum 4 have treatment factors applied to them.

We show 
three stratum-by-stratum designs, using different criteria: $D_S$, $(DP)_S$ and compound criteria with weights given as $\kappa_{DP}=\kappa_L=\kappa_{DF}=1/3$. The construction steps for the stratum-by-stratum approach, again assuming three-level factorials as the candidate set, were:
\begin{enumerate}
\item Generate a random non-singular design for $X_1$ and $X_2$, say $\mathfrak{X}_1^*$, using $m_1=n_1=10$ units. The model includes $p_1=6$ parameters, four main effects (linear and quadratic), one linear-by-linear interaction effect, and the intercept. \label{stepEx3:phase1.1}
\item Optimize, according to the criterion function chosen, the design $\mathfrak{X}_1^*$ in (\ref{stepEx3:phase1.1}) by performing point exchanges.  For inference-based criteria, treatment labels are assigned to the rows of $\mathfrak{X}_1^*$ to calculate PE degrees of freedom, where the implicit model is based on the full treatment effects matrix $\mathbf{T}_1$. For inference-based criteria, we use both the treatment-based and the polynomial-based models to optimize the design. Otherwise, we use just the polynomial-based model matrix $\mathbf{X}_1$ for $p_1-1$ parameters. Let $\mathfrak{X}_1^*$ be the optimized factor level combinations found that will be randomized to oven runs or whole-plots (stratum 1). \label{stepEx3:phase1.2}
\item Each row in $\mathfrak{X}_1^*$, from step (\ref{stepEx3:phase1.2}), is replicated three times with each replicate being assigned to one of the $n_2=b=3$ batches of size 10, to form a blocked design. Let $\mathfrak{X}_1^{**}$ be this enlarged matrix ($m_3=30$ rows).\label{stepEx3:phase2.1}
\item The rows of the matrix $\mathfrak{X}_1^{**}$ from step (\ref{stepEx3:phase2.1}) are replicated twice in order to form $b=30$ blocks of size two ($n_3=2$). Let $\mathfrak{X}_1$ be this enlarged matrix with $m_4=60$ rows.  \label{stepEx3:phase2.2}
\item With $\mathfrak{X}_1$ fixed, generate a non-singular random blocked initial design for $X_3,~X_4,~X_5$. Let $\mathfrak{X}_4$ be the factor-level combinations in this stratum 4. Note that the combinations of whole-plots and blocks from step (\ref{stepEx3:phase2.2}) act as new blocks. Thus the number of blocks is $b=30$, each of size two. The approximating model is block effects plus linear and quadratic effects of $X_3,~X_4,~X_5$, and linear-by-linear interaction terms between all factors except $X_1\times X_2$. This polynomial model has $p_4=16$ parameters such that, the corresponding $\mathbf{X}_4$ matrix has $15$ columns.\label{stepEx3:phase3.1}
\item With $\mathfrak{X}_1$ fixed, optimize $\mathfrak{X}_4$ in (\ref{stepEx3:phase3.1}), according to the criterion function chosen, by performing point exchanges in $\mathfrak{X}_4$. For PE degrees of freedom, the model has block effects and $\mathbf{T}_4$, the full treatment effects matrix based on $\mathfrak{X}_4$ plus their interactions with treatments based on 
 $\mathfrak{X}_1$.\label{stepEx3:phase3.2}
\item As the three original blocks (from step (\ref{stepEx3:phase2.1})) are replicates of whole-plot treatments, optimize further the design found in (\ref{stepEx3:phase3.2}) by swapping combinations of whole-plots among blocks restricting to whole-plots with the same levels of $X_1$ and $X_2$ ($\mathfrak{X}_1$). Thus a constrained interchange algorithm is applied in which the blocking system is $b=3$ blocks of size 20 and the approximating model is the full second-order model with $p=21$. The final design is the optimum design given the initial design.
\item Repeat, from step \ref{stepEx3:phase1.1}, for several initial designs. The best design found among the repetitions is declared to be the optimum.
\end{enumerate}

The designs found are shown in Tables \ref{tab:Ex3D}-\ref{tab:Ex3CP}. For the stratum-by-stratum approach, we used at least 2,000 initial designs. Even using 20,000 initial designs, JMP returned a design that was inferior to our MSS$_{D_S}$ design (95\% efficiency), so we believe the MSS$_{D_S}$ design is a fixed effects $D$-optimal design and we do not show the JMP design. The MSS$_{D_S}$ design allows only one PE degree of freedom in each stratum, as shown in Table \ref{tab:dfEx3}, with both this and the compromise design allowing for model lack-of-fit in both strata. The MSS$_{(DP)_S}$ design allows 3 and 12 PE degrees of freedom in the whole-plot and sub-plot strata, respectively, but no model lack-of-fit in the whole plot stratum can be checked. We note that, during the construction of the design, it allowed 4 PE degrees of freedom in the whole-plot stratum, however, one ended up being confounded with treatment effects in the sub-plot stratum. The compromise design gives 8 degrees of freedom for pure error, seven for lack-of-fit in the lower stratum and three degrees of freedom for PE in the whole-plot stratum. It also allows one higher-order term to be fitted in the whole-plot stratum if needed. 

\begin{table}
\caption{MSS$_{D_S}$ design for Example 3, a strip-split-plot structure (Ovens(10)*Batches(3))/Runs(2), with two HS and three ES 3-level factors}\label{tab:Ex3D}
\centering
\setlength\tabcolsep{0.15cm}
\renewcommand{\arraystretch}{.8}
\small{
\begin{tabular}{rcrrr|rrr|rrr}\hline
&&\multicolumn{9}{c}{Batch}\\\cline{3-11}
\multicolumn{1}{c}{Oven}& &\multicolumn{3}{c}{1}&\multicolumn{3}{c}{2}&\multicolumn{3}{c}{3}\\
$\begin{array}{cc}X_1&X_2\end{array}$&&$X_3$&$X_4$&$X_5$&$X_3$&$X_4$&$X_5$&$X_3$&$X_4$&$X_5$\\\hline
\multirow{2}{*}{$\begin{array}{rr}\mbox{\normalsize -1}&\mbox{\normalsize -1}\end{array}$}
&&-1 & -1 & -1 & -1 & 1 & -1 & -1 & 1 & 1 \\ 
&& 1 & 1 & -1 & 1 & 1 & 1 & 1 & -1 & -1 \\ 
\hline
\multirow{2}{*}{$\begin{array}{rr}\mbox{\normalsize -1}&\mbox{\normalsize ~0}\end{array}$}
&& 1 & 1 & 1 & 0 & 0 & 1 & -1 & -1 & -1 \\ 
&& 1 & -1 & -1 & 1 & 1 & -1 & 1 & -1 & 1 \\ 
\hline
\multirow{2}{*}{$\begin{array}{rr}\mbox{\normalsize -1}&\mbox{\normalsize ~1}\end{array}$}
&&  -1 & -1 & 1 & -1 & -1 & -1 & -1 & 1 & -1 \\ 
&&  1 & 1 & 0 & -1 & 1 & 1 & 1 & -1 & 1 \\ 
\hline
\multirow{2}{*}{$\begin{array}{rr}\mbox{\normalsize ~0}&\mbox{\normalsize -1}\end{array}$}
&&  0 & 0 & -1 & -1 & 1 & -1 & -1 & -1 & 1 \\ 
&&  1 & -1 & 1 & 0 & -1 & 0 & 0 & 1 & 0 \\ 
\hline
\multirow{2}{*}{$\begin{array}{rr}\mbox{\normalsize ~0}&\mbox{\normalsize ~0}\end{array}$}
&&  0 & 1 & -1 & 0 & 0 & -1 & 1 & 0 & 0 \\ 
&&  -1 & 0 & 0 & -1 & 1 & 0 & 0 & 1 & -1 \\ 
\hline
\multirow{2}{*}{$\begin{array}{rr}\mbox{\normalsize ~0}&\mbox{\normalsize ~1}\end{array}$}
&&  0 & 0 & 1 & 1 & 1 & 1 & -1 & -1 & 0 \\ 
&&  1 & -1 & -1 & 0 & -1 & 0 & 1 & 0 & -1 \\ 
\hline
\multirow{2}{*}{$\begin{array}{rr}\mbox{\normalsize ~1}&\mbox{\normalsize -1}\end{array}$}
&&  1 & 1 & -1 & -1 & -1 & -1 & -1 & 1 & 1 \\ 
&&  0 & -1 & 1 & 1 & 0 & 1 & 1 & -1 & -1 \\ 
\hline
\multirow{2}{*}{$\begin{array}{rr}\mbox{\normalsize ~1}&\mbox{\normalsize ~0}\end{array}$}
&&  1 & 1 & 1 & -1 & -1 & 1 & 0 & 1 & 1 \\ 
&&  -1 & 0 & 0 & 1 & -1 & -1 & -1 & 0 & -1 \\
\hline
\multirow{2}{*}{$\begin{array}{rr}\mbox{\normalsize ~1}&\mbox{\normalsize ~1}\end{array}$} 
&&  1 & -1 & 1 & -1 & 1 & 1 & -1 & -1 & -1 \\ 
&& -1 & 1 & 1 & 1 & 1 & -1 & 1 & 1 & 0 \\ 
\hline
\multirow{2}{*}{$\begin{array}{rr}\mbox{\normalsize ~1}&\mbox{\normalsize ~1}\end{array}$}
&&  -1 & 1 & -1 & -1 & -1 & 1 & 1 & -1 & 1 \\ 
&&  1 & -1 & -1 & -1 & 1 & -1 & 1 & 1 & -1 \\
   \hline
\end{tabular}}
\end{table}

\begin{table}
\caption{MSS$_{(DP)_S}$ design for Example 3, a strip-split-plot structure (Ovens(10)*Batches(3))/Runs(2), with two HS and three ES 3-level factors}\label{tab:Ex3DP}
\centering
\setlength\tabcolsep{0.15cm}
\renewcommand{\arraystretch}{.8}
\small{
\begin{tabular}{rcrrr|rrr|rrr}\hline
&&\multicolumn{9}{c}{Batch}\\\cline{3-11}
\multicolumn{1}{c}{Oven}& &\multicolumn{3}{c}{1}&\multicolumn{3}{c}{2}&\multicolumn{3}{c}{3}\\
$\begin{array}{cc}X_1&X_2\end{array}$&&$X_3$&$X_4$&$X_5$&$X_3$&$X_4$&$X_5$&$X_3$&$X_4$&$X_5$\\\hline
\multirow{2}{*}{$\begin{array}{rr}\mbox{\normalsize -1}&\mbox{\normalsize -1}\end{array}$}
&&-1 & 1 & -1 & -1 & 0 & 1 & 1 & -1 & 0 \\ 
&&  1 & 0 & 1 & 1 & 1 & -1 & 1 & 1 & -1 \\ 
\hline
\multirow{2}{*}{$\begin{array}{rr}\mbox{\normalsize -1}&\mbox{\normalsize -1}\end{array}$}
&&-1 & 0 & 1 & 1 & 1 & 1 & -1 & 1 & -1 \\ 
&&  1 & -1 & 0 & 0 & -1 & -1 & 1 & 0 & 1 \\  
\hline
\multirow{2}{*}{$\begin{array}{rr}\mbox{\normalsize -1}&\mbox{\normalsize ~0}\end{array}$}
&&  -1 & 1 & 0 & 1 & 0 & -1 & -1 & -1 & 1 \\ 
&&  0 & -1 & 1 & 0 & 1 & 1 & 0 & 0 & -1 \\ 
\hline
\multirow{2}{*}{$\begin{array}{rr}\mbox{\normalsize -1}&\mbox{\normalsize ~0}\end{array}$}
&&1 & 0 & -1 & -1 & -1 & 1 & 0 & 0 & -1 \\ 
&&  0 & 1 & 1 & 0 & 0 & -1 & -1 & -1 & 1 \\ 
\hline
\multirow{2}{*}{$\begin{array}{rr}\mbox{\normalsize -1}&\mbox{\normalsize ~1}\end{array}$}
&& 1 & 1 & -1 & 1 & -1 & 1 & 1 & 1 & -1 \\ 
&&  -1 & -1 & -1 & -1 & -1 & -1 & 1 & -1 & 1 \\ 
\hline
\multirow{2}{*}{$\begin{array}{rr}\mbox{\normalsize ~0}&\mbox{\normalsize ~1}\end{array}$}
&&  -1 & -1 & -1 & 1 & 1 & 1 & -1 & 1 & 1 \\ 
&&  -1 & 1 & 1 & -1 & 1 & -1 & -1 & -1 & -1 \\ 
\hline
\multirow{2}{*}{$\begin{array}{rr}\mbox{\normalsize ~0}&\mbox{\normalsize ~1}\end{array}$}
&&    1 & -1 & -1 & 1 & -1 & -1 & 1 & -1 & -1 \\ 
&&  1 & 0 & 0 & 1 & 0 & 0 & 1 & 0 & 0 \\ 
\hline
\multirow{2}{*}{$\begin{array}{rr}\mbox{\normalsize ~1}&\mbox{\normalsize -1}\end{array}$}
&& 1 & 0 & -1 & -1 & 1 & 0 & 0 & 1 & 1 \\ 
&&  -1 & -1 & 1 & 1 & -1 & 1 & -1 & -1 & -1 \\ 
\hline
\multirow{2}{*}{$\begin{array}{rr}\mbox{\normalsize ~1}&\mbox{\normalsize ~1}\end{array}$}
&& -1 & 1 & 1 & -1 & 1 & 1 & 1 & 1 & 1 \\ 
&&  1 & 1 & -1 & 1 & 1 & -1 & 0 & -1 & 0 \\ 
\hline
\multirow{2}{*}{$\begin{array}{rr}\mbox{\normalsize ~1}&\mbox{\normalsize ~1}\end{array}$}
&& 0 & -1 & 0 & -1 & 1 & -1 & 1 & -1 & 0 \\ 
&&  1 & 1 & 1 & 1 & -1 & 0 & -1 & 1 & -1 \\ 
   \hline
\end{tabular}}
\end{table}

\begin{table}
\caption{MSS$_{CP_{\boldsymbol{\kappa}}}$ design for Example 3, a strip-split-plot structure (Ovens(10)*Batches(3))/Runs(2), with two HS and three ES 3-level factors}\label{tab:Ex3CP}\centering
\setlength\tabcolsep{0.15cm}
\renewcommand{\arraystretch}{.8}
\small{
\begin{tabular}{rcrrr|rrr|rrr}\hline
&&\multicolumn{9}{c}{Batch}\\\cline{3-11}
\multicolumn{1}{c}{Oven}& &\multicolumn{3}{c}{1}&\multicolumn{3}{c}{2}&\multicolumn{3}{c}{3}\\
$\begin{array}{cc}X_1&X_2\end{array}$&&$X_3$&$X_4$&$X_5$&$X_3$&$X_4$&$X_5$&$X_3$&$X_4$&$X_5$\\\hline
\multirow{2}{*}{$\begin{array}{rr}\mbox{\normalsize -1}&\mbox{\normalsize -1}\end{array}$}
&&-1 & -1 & -1 & -1 & -1 & 1 & 1 & 1 & -1 \\ 
&&  1 & -1 & 1 & 1 & 1 & 1 & 1 & -1 & 1 \\ 
\hline
\multirow{2}{*}{$\begin{array}{rr}\mbox{\normalsize -1}&\mbox{\normalsize -1}\end{array}$}
&&  -1 & 1 & 1 & -1 & -1 & -1 & 1 & -1 & -1 \\ 
&&  -1 & -1 & -1 & 1 & 1 & -1 & -1 & 0 & 1 \\ 
\hline
\multirow{2}{*}{$\begin{array}{rr}\mbox{\normalsize -1}&\mbox{\normalsize ~0}\end{array}$}
&&  1 & 0 & 0 & -1 & -1 & 1 & -1 & -1 & 1 \\ 
&&  -1 & 1 & 1 & -1 & 1 & -1 & -1 & 1 & -1 \\ 
\hline
\multirow{2}{*}{$\begin{array}{rr}\mbox{\normalsize -1}&\mbox{\normalsize ~1}\end{array}$} 
&&  -1 & 1 & -1 & 1 & -1 & -1 & -1 & 1 & -1 \\ 
&&  1 & -1 & -1 & 1 & 1 & 1 & 1 & 1 & 1 \\ 
\hline
\multirow{2}{*}{$\begin{array}{rr}\mbox{\normalsize ~0}&\mbox{\normalsize ~0}\end{array}$}
&&  -1 & 1 & 0 & 0 & 0 & 1 & 0 & 0 & -1 \\ 
&&  0 & 0 & -1 & -1 & -1 & 0 & -1 & 1 & 0 \\ 
\hline
\multirow{2}{*}{$\begin{array}{rr}\mbox{\normalsize ~0}&\mbox{\normalsize ~1}\end{array}$}
&& 1 & 1 & -1 & 1 & -1 & -1 & -1 & 0 & 0 \\ 
&&  -1 & 0 & 0 & -1 & -1 & 1 & 1 & 1 & -1 \\ 
\hline
\multirow{2}{*}{$\begin{array}{rr}\mbox{\normalsize ~1}&\mbox{\normalsize -1}\end{array}$}
&&1 & 0 & -1 & 1 & 1 & 1 & -1 & 0 & 1 \\ 
&&  0 & -1 & 1 & -1 & 1 & -1 & 0 & 1 & -1 \\ 
\hline
\multirow{2}{*}{$\begin{array}{rr}\mbox{\normalsize ~1}&\mbox{\normalsize -1}\end{array}$}
&& 0 & 1 & -1 & 1 & 1 & 1 & 1 & -1 & -1 \\ 
&&  -1 & 0 & 1 & 1 & -1 & -1 & -1 & 1 & -1 \\ 
\hline
\multirow{2}{*}{$\begin{array}{rr}\mbox{\normalsize ~1}&\mbox{\normalsize ~1}\end{array}$}
&&  0 & -1 & 1 & 1 & -1 & 1 & 0 & -1 & 0 \\ 
&&  1 & 1 & -1 & 0 & 1 & 0 & -1 & 1 & 1 \\ 
\hline
\multirow{2}{*}{$\begin{array}{rr}\mbox{\normalsize ~1}&\mbox{\normalsize ~1}\end{array}$}
&& 0 & 1 & 1 & 1 & -1 & 1 & -1 & -1 & -1 \\ 
&&  -1 & -1 & -1 & 0 & 1 & 0 & 1 & 0 & 0 \\ 
   \hline
\end{tabular}}
\end{table}

\begin{table}
\caption{Skeleton ANOVA of designs for Example 3, a strip-split-plot structure (Ovens(10)*Batches(3))/Runs(2), with two HS and three ES 3-level factors}\label{tab:dfEx3}
\centering
\setlength\tabcolsep{0.15cm}
\renewcommand{\arraystretch}{.8}
\begin{tabular}{llrrr}
\hline
& &\multicolumn{3}{c}{Designs}\\\cline{3-5}
    Stratum   &Source                                   &\multicolumn{1}{c}{MSS$_{D_S}$}&\multicolumn{1}{c}{MSS$_{(DP)_S}$}&\multicolumn{1}{c}{MSS$_{CP_{\boldsymbol{\kappa}}}$}\\\hline
Ovens&\multicolumn{1}{l}{Treat ($\mathbf{T}_1$):}                     &8&5&6\\
  &\multicolumn{1}{l}{~~\textit{2$^{nd}$ order}}              &\textit{5}    &\textit{5}        &\textit{5} \\
  &\multicolumn{1}{l}{~~\textit{Lack-of-Fit}}                 &\textit{3}    &\textit{0}       &\textit{1}\\
  & \multicolumn{1}{l}{Treat ($\mathbf{T}_4$): Lack-of-Fit} & 0 & 1 & 0\\
  &\multicolumn{1}{l}{PE}                            &1&3&3\\ \cline{2-5}  
  &Total&9&9&9\\\hline
Batches&					&2&2     &2\\\hline
Ovens*Batches& \multicolumn{1}{l}{Treat ($\mathbf{T}_4$): Lack-of-Fit} & {17} &{9}  & {10}\\
& \multicolumn{1}{l}{PE} & 1& 9 & 8\\ \cline{2-5} 
& Total &18&18   &18\\\hline
Runs     &\multicolumn{1}{l}{Treat ($\mathbf{T}_4$):}          &29&18&22\\
                &\multicolumn{1}{l}{~~\textit{2$^{nd}$ order}}&\textit{15}    &\textit{15}       &\textit{15} \\
                &\multicolumn{1}{l}{~~\textit{Lack-of-Fit}}   &\textit{14}    &\textit{3}       &\textit{7}\\
                &\multicolumn{1}{l}{PE}              &1&12&8\\ \cline{2-5}
& \multicolumn{1}{l}{Total} & 30 & 30 & 30\\ \hline
\multicolumn{2}{l}{Total}                            &59&59&59\\\hline
\end{tabular}
\end{table}

The efficiencies based on the mixed model variance and covariance matrix for the fixed-effect estimators are shown in Table \ref{tab:effEx3}, assuming different values for the variance component ratios. They are calculated based on the usual $D_S$ and $A_S$ criteria and the reference is the design that showed the best performance for the fixed effects $D$ criterion, as already mentioned, the MSS$_{D_S}$ design. The compromise design performs quite well, compared to MSS$_{D_S}$, it loses very little in terms of efficiency while allowing PE degrees of freedom in both strata.

\begin{table}
\caption{$D_S$- and $A_S$-efficiencies, relative to the best fixed-effects $D$ design obtained, for the strip-split-plot designs for Example 3}\label{tab:effEx3}
\centering
\setlength\tabcolsep{0.15cm}
\renewcommand{\arraystretch}{.8}
\begin{tabular}{ccccrrr}\hline
& & & &\multicolumn{3}{c}{Designs}\\\cline{5-6}
Criterion&$\eta_{Oven}$&$\eta_{Batch}$&$\eta_{Oven* Batch}$&MSS$_{(DP)_S}$&MSS$_{CP_{\boldsymbol{\kappa}}}$\\\hline
$D_S$&  1&  1&  1&	 89.21~~~~&	 98.54~~	\\
     &100&  1&  1&	88.80~~~~&	98.39~~	\\
     &100&100&  1&	88.81~~~~&	98.33~~	\\
     &100&100&100& 89.00~~~~&	99.49~~	\\\hline
$A_S$&  1&  1&  1&	76.79~~~~&91.26~~	\\
     &100&  1&  1&76.04~~~~&89.99~~	\\
     &100&100&  1&	76.04~~~~&	89.99~~	\\
     &100&100&100&	76.03~~~~&	90.00~~	\\\hline
\end{tabular}
\end{table}

\subsection{Example 4: polypropylene coating experiment in a split-row$\times$split-column design}

\begin{figure}
    \centering
    \includegraphics[width=12cm, height=10cm]{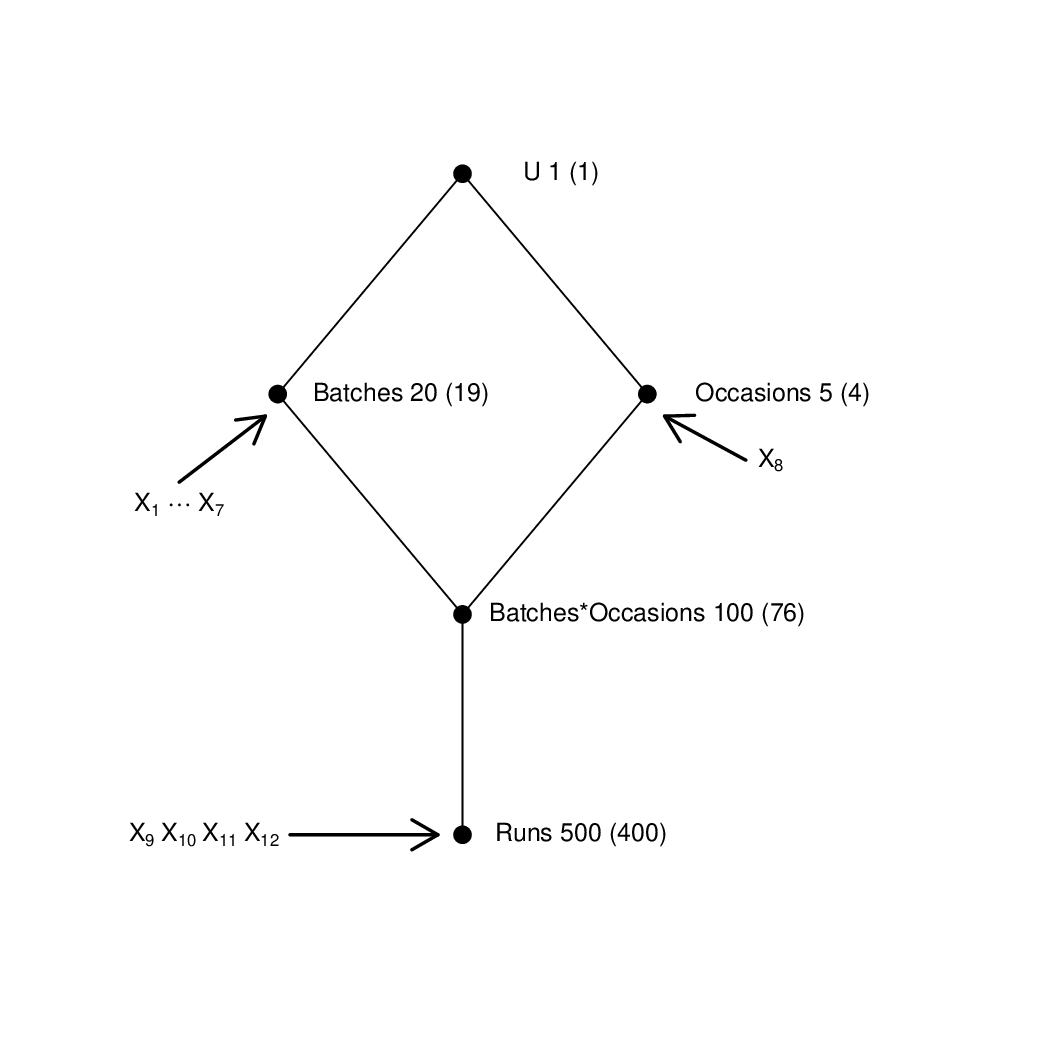}\vspace{-1cm}
    \caption{Hasse diagram of the unit structure of the split-row$\times$split-column design for Example 4.}
    \label{fig:hasseEx4}
\end{figure}

In discussing the analysis of a data set on polypropylene adhesion coatings, \cite{GoosGilmour2012} noted that the design of the experiment had several weaknesses and suggested an improved structure for the design, without giving any details of how such a design could be constructed. The structure they proposed had 20 batches, with combinations of levels of seven two-level factors applied to them, each used on each of five occasions, with five types of coating applied to them, and with five runs in each Batch-by-Occasion combination, as in the experiment performed. The improvement suggested was that the levels of the four remaining three-level factors should be applied to runs within the Batch-by-Occasion combinations. This defines the unit structure as (Batches*Occasions)/Runs. There were also three measurements made within each run, but these do not affect the design problem. Ideally, the design would be better if more than five occasions could be used, however, that might be prohibitive in practice since the design is already quite large. Using our notation, the design structure is represented by Batches(20)*Occasions(5)/Runs(5). The Hasse diagram of the unit structure of the design, including the factors to be applied to the units in each stratum, is given in Figure \ref{fig:hasseEx4}. There are four strata, with strata 1, 2 and 4 each having a set of treatment factors applied.

The steps for the stratum-by-stratum design construction, similar to Example 3, are:
\begin{enumerate}
\item Generate a random non-singular design for $X_1, X_2,\ldots,X_7$ in $m_1=n_1=20$ runs (Batches). Given the knowledge of the factor's relations, the combination of $X_3$ and $X_4$ both at their highest levels was to be avoided and the polynomial model, in this stratum, would include only linear effects of the seven factors and the linear-by-linear interaction terms involving $X_1$. Thus, the candidate set to be used is the two-level factorial with the aforementioned restriction. The model includes $p_1=14$ parameters. Let $\mathfrak{X}_1^*$ be the matrix of factor level combinations generated.
\label{stepEx4:step1} 
\item Optimize, according to the criterion function chosen, the design $\mathfrak{X}_1^*$ in (\ref{stepEx4:step1}) by performing point exchanges.  For inference-based criteria, treatment labels are assigned to the rows of $\mathfrak{X}_1^*$ to calculate PE degrees of freedom, where the implicit model is based on the full treatment effects matrix $\mathbf{T}_1$. For inference-based criteria, we use both models to optimize the design, the treatment-based and the polynomial-based. Otherwise, we use just the polynomial-based model matrix, $\mathbf{X}_1$ with $13$ columns (stratum 1). Let $\mathfrak{X}_1^*$ be the optimized factor level combinations found. The treatments from this matrix will be randomized to Batches or whole-plots. \label{stepEx4:step2}
\item Find an unblocked design for $X_8$ in $m_2=n_2=5$ runs (Occasions) in stratum 2. In this case, as the number of runs is the same as treatments, no optimization is required, just an allocation of each treatment to each run. Let $\mathfrak{X}_2$ be the matrix representing the allocation.\label{stepEx4:step3}
\item Combine the designs in (\ref{stepEx4:step2}) and (\ref{stepEx4:step3}) by replicating each row of $\mathfrak{X}_1$ five times ($n_2=5$) and binding them to $\mathfrak{X}_2$. Let $\mathfrak{X}_3^*$ be the bound matrix, with $m_3=100$ rows, giving the resulting factor level combinations in this step (stratum 3).
\item Further replicate each row of $\mathfrak{X}_3^*$ five times ($n_4=5$), each group of five denoted as a block ($b=20\times5=100$ blocks each of size five). Let $\mathfrak{X}_3$ be the enlarged matrix with $m_4=500$ rows.\label{stepEx4:step4}
\item Generate a random blocked design for $X_9,\ldots X_{12}$ in $b=100$ blocks of size $5$, combined with $\mathfrak{X}_3$ above. Let $\mathfrak{X}_4$ be the level combinations for $X_9,\ldots X_{12}$. The polynomial model in this step is quite large, $p_4=261$, including linear, quadratic and linear-by-linear effects for the new factors, two-factor interactions involving these factors and the ones applied at previous strata and some three-factor interactions involving $X_8$ (for a detailed description of the terms in the model, see \cite{GoosGilmour2012}). \label{stepEx4:step5}
\item With $\mathfrak{X}_3$ fixed, optimize $\mathfrak{X}_4$ for the blocking scheme specified. Depending on the criterion used, the polynomial and the treatment model, using $\mathbf{T}_4$, might be required for the optimization. \label{stepEx4:step6}
\item If there are replications in the row and/or column designs, search for swaps among 
	blocks that improve efficiency at higher strata.
\item Repeat, from (\ref{stepEx3:phase1.1}), for several initial designs. The best design found among the repetitions is declared to be the optimum.
\end{enumerate}

The most challenging aspect of the design construction for this problem is the optimization of the design in step (\ref{stepEx4:step6}) since the design has a size of 500 runs and the polynomial model includes 261 parameters. So we constructed designs using only two starting designs due to the computer time required, even when the code allowed for parallel program running.

We constructed three designs for the problem, MSS$_{D_S}$, MSS$_{(DP)_S}$ and MSS$_{CP_{\boldsymbol{\kappa}}}$. 
Because they are huge, they are stored in a zip file (Supplementary Material). Unfortunately, we do not have a baseline to compare our designs with, since we were unable to get JMP to find a fixed effects $D$-optimal design. In Table 
\ref{tab:dfEx4} we show the degrees of freedom breakdown and in Table \ref{tab:effEx4} the efficiencies of the designs relative to the MSS$_{D_S}$ design. Once again, the MSS$_{D_S}$ does not provide PE degrees of freedom in any stratum. The $MSS_{(DP)_S}$ allows 3, 3 and 77 PE degrees of freedom in the Batches, Batches$*$Occasions and Runs strata, respectively. The MSS$_{CP}$ design allows 2, 1 and 39 PE degrees of freedom in the Batches, Batches$*$Occasions and Runs strata, respectively. Similarly to Example 3, in both designs, three PE degrees of freedom in the Batches stratum were lost to treatment effects from the Runs stratum. Table \ref{tab:effEx4} shows that in terms of $D_S$-efficiency, both designs are robust to changes in the variance components.

\begin{table}
\caption{Skeleton ANOVA of three designs for Example 4 (Batches($20$)$*$Occasions($5$)/Runs($5$))}
\label{tab:dfEx4}
 \centering
\setlength\tabcolsep{0.15cm}
\renewcommand{\arraystretch}{.8}
\begin{tabular}{llrrr}\hline
&&\multicolumn{3}{c}{Designs}\\ 
\cline{3-5}
Stratum&{Source}&MSS$_{D_S}$&MSS$_{(DP)_S}$&MSS$_{CP_{\boldsymbol{\kappa}}}$\\	\hline
{Batches}&{Treat ($\mathbf{T}_1$):}&{19}&{13}&{14}\\
&\multicolumn{1}{l}{~~\textit{Model}}&\textit{13}&\textit{13}&\textit{13}\\
&\multicolumn{1}{l}{~~\textit{Lack-of-Fit}}& \textit{6}&\textit{0}& \textit{1}\\
& Treat ($\mathbf{T}_4$): Lack-of-Fit & {0}& {3}& 3\\
& PE&0&3&2\\ \cline{2-5}
&Total&19&19&19\\\hline
{Occasions}&Treat ($\mathbf{T}_2$):&4&4&4\\
&\multicolumn{1}{l}{~~\textit{Model}}&\textit{4}&\textit{4}&\textit{4}\\
&PE&0&0&0\\ \cline{2-5}
&Total&4&4&4\\\hline
{Batches*Occasions}&Treat ($\mathbf{T}_3$): &76&52&56\\
&\multicolumn{1}{l}{~~\textit{Model}}&\textit{52}&\textit{52}&\textit{52}\\
&\multicolumn{1}{l}{~~\textit{Lack-of-Fit}}&\textit{24}&\textit{0}&\textit{4}\\
&Treat ($\mathbf{T}_4$): Lack-of-Fit & 0& 21 & 19\\
&PE&0&3&1\\ \cline{2-5}
&Total&76&76&76\\\hline
{Runs}&Treat ($\mathbf{T}_4$):&400&323&361\\
&\multicolumn{1}{l}{~~\textit{Model}}&\textit{260}&\textit{260}&\textit{260}\\
&\multicolumn{1}{l}{~~\textit{Lack-of-Fit}}&\textit{140}&\textit{63}&\textit{101}\\
&PE&0&77&39\\ \cline{2-5}
&Total&400&400&400\\ \hline
Total & & 499& 499 & 499\\\hline
		\end{tabular}
  \end{table}

\begin{table}
\caption{$D_S$- and $A_S$-efficiencies of designs for Example 4, relative to the MSS$_D$ optimum design}\label{tab:effEx4}
\centering
\setlength\tabcolsep{0.15cm}
\renewcommand{\arraystretch}{.6}
\begin{tabular}{cccrrcrr}\hline
&&&\multicolumn{5}{c}{Criterion}\\\cline{4-8}
&&&\multicolumn{2}{c}{$D_S$}&&\multicolumn{2}{c}{$A_S$}\\
&&&\multicolumn{2}{c}{Designs}&&\multicolumn{2}{c}{Designs}\\\cline{4-5}\cline{7-8}
$\eta_{Batch}$ & $\eta_{Occ}$ & $\eta_{Batch\star Occ}$ & MSS$_{(DP)_S}$ & MSS$_{CP_{\boldsymbol{\kappa}}}$&&MSS$_{(DP)_S}$ & MSS$_{CP_{\boldsymbol{\kappa}}}$\\ 
  \hline
1 &  1 &  1 & 86.11 & 90.96 && 75.42 & 96.44 \\ 
   10 &  1 &  1 & 86.10 & 90.95 && 75.78 & 95.75 \\ 
   100 &  1 &  1 & 86.10 & 90.95 && 77.21 & 93.32 \\ 
   1 &  10 &  1 & 86.11 & 90.96 && 88.50 & 98.55 \\ 
   10 &  10 &  1 & 86.10 & 90.95 && 87.83 & 98.11 \\ 
   100 &  10 &  1 & 86.10 & 90.95 && 84.07 & 95.61 \\ 
   1 &  100 &  1 & 86.11 & 90.96 && 98.18 & 99.79 \\ 
   10 &  100 &  1 & 86.10 & 90.95 && 97.96 & 99.71 \\ 
   100 &  100 &  1 & 86.10 & 90.95 && 96.03 & 99.01 \\ 
   1 &  1 &  10 & 85.80 & 90.73 && 77.63 & 93.03 \\ 
   10 &  1 &  10 & 85.79 & 90.72 && 77.67 & 92.94 \\ 
   100 &  1 &  10 & 85.79 & 90.72 && 77.90 & 92.44 \\ 
   1 &  10 &  10 & 85.80 & 90.73 && 84.69 & 95.51 \\ 
   10 &  10 &  10 & 85.79 & 90.72 && 84.43 & 95.36 \\ 
   100 &  10 &  10 & 85.79 & 90.72 && 82.69 & 94.31 \\ 
   1 &  100 &  10 & 85.80 & 90.73 && 96.31 & 99.01 \\ 
   10 &  100 &  10 & 85.79 & 90.72 && 96.14 & 98.95 \\ 
   100 &  100 &  10 & 85.79 & 90.72 && 94.53 & 98.36 \\ 
   1 &  1 &  100 & 85.75 & 90.69 && 79.16 & 91.30 \\ 
   10 &  1 &  100 & 85.75 & 90.69 && 79.15 & 91.30 \\ 
   100 &  1 &  100 & 85.75 & 90.69 && 79.08 & 91.32 \\ 
   1 &  10 &  100 & 85.75 & 90.69 && 80.46 & 91.92 \\ 
   10 &  10 &  100 & 85.75 & 90.69 && 80.44 & 91.92 \\ 
   100 &  10 &  100 & 85.75 & 90.69 && 80.28 & 91.89 \\ 
   1 &  100 &  100 & 85.75 & 90.69 && 87.97 & 95.28 \\ 
   10 &  100 &  100 & 85.75 & 90.69 && 87.93 & 95.27 \\ 
   100 &  100 &  100 & 85.75 & 90.69 && 87.47 & 95.10 \\ 
   \hline
\end{tabular}
\end{table}

\section{Discussion}\label{sec:discussion}

The main contribution of this paper is to show how good designs can be found for any multi-stratum structure consisting of nested and crossed blocking factors which are defined by restrictions in the randomization. The stratum-by-stratum approach to construction has the advantage that it generalizes to any structure without the computational requirements exploding. In this paper, we have used compound design optimality criteria, but the general strategy could be used with any criterion. Exploring how the designs change with different requirements is an interesting area for future research.

The compound criteria used in this work give a reasonable compromise between point estimation of parameters, inference on parameters and allowing for lack of fit. As such, they allow designs with many of the same properties as classical designs, such as central composite designs, to be found in even the most complex of design structures.

Previous work which used the stratum-by-stratum approach and compound criteria (\cite{GilmourTrinca2012}, \cite{TrincaGilmour2017}) can be considered as special cases of the complex structures studied here, so that the present paper can be considered as the most complete solution to date of how to design response surface experiments.






\section*{Supplementary Materials}
\begin{description}
    \item[designs$\_$Example4.zip:] designs for Example 4.
\end{description}

\section*{Acknowledgments}
The first author gratefully acknowledges financial support from FAPESP grant 14/01818-0. The second author gratefully acknowledges financial support from EPSRC grant EP/T021624/1.

\section*{Disclosure Statement}
The authors report there are no competing interests to declare.

\bibliography{Bibliography}

\begin{thebibliography}{}

\bibitem[Borrotti et~al., 2023]{Borrotietal2023}
Borrotti, M., Sambo, F., and Mylona, K. (2023).
\newblock Multi-objective optimisation of split-plot experiments.
\newblock {\em Econometrics and Statistics}, 28:163--172.

\bibitem[Borrotti et~al., 2017]{Borrotietal2017}
Borrotti, M., Sambo, F., Mylona, K., and Gilmour, S.~G. (2017).
\newblock A multi-objective coordinate-exchange two-phase local search
  algorithm for multi-stratum experiments.
\newblock {\em Statistics and Computing}, 27:469--481.

\bibitem[Cao et~al., 2017]{Caoetal2017}
Cao, Y., Wulff, S.~S., and Robinson, T.~J. (2017).
\newblock {DP}-optimality in terms of multiple criteria and its application to
  the split-plot design.
\newblock {\em Journal of Quality Technology}, 49:27--45.

\bibitem[Cuervo et~al., 2017]{Cuervoetal2017}
Cuervo, D.~P., Goos, P., and S\"orensen, K. (2017).
\newblock An algorithmic framework for generating optimal two-stratum
  experimental designs.
\newblock {\em Computational Statistics and Data Analysis}, 115:224--249.

\bibitem[Gilmour and Trinca, 2003]{GilmourTrinca2003}
Gilmour, S. and Trinca, L. (2003).
\newblock Row-column response surface designs.
\newblock {\em Journal of Quality Technology}, 35.

\bibitem[Gilmour and Trinca, 2012]{GilmourTrinca2012}
Gilmour, S. and Trinca, L. (2012).
\newblock Optimum design of experiments for statistical inference.
\newblock {\em Journal of the Royal Statistical Society. Series C: Applied
  Statistics}, 61.

\bibitem[Goos, 2002]{Goos2002}
Goos, P. (2002).
\newblock {\em The Optimal Design of Blocked and Split-Plot Experiments}.
\newblock New York: Springer.

\bibitem[Goos, 2006]{Goos2006}
Goos, P. (2006).
\newblock Optimal versus orthogonal and equivalent-estimation design of blocked
  and split-plot experiments.
\newblock {\em Statistica Neerlandica}, 60:361--378.

\bibitem[Goos and Gilmour, 2012]{GoosGilmour2012}
Goos, P. and Gilmour, S.~G. (2012).
\newblock A general strategy for analyzing data from split-plot and
  multistratum experimental designs.
\newblock {\em Technometrics}, 54:340--354.

\bibitem[Goos and Vandebroek, 2003]{GoosVandebroek2003}
Goos, P. and Vandebroek, M. (2003).
\newblock D-optimal split-plot designs with given numbers and sizes of whole
  plots.
\newblock {\em Technometrics}, 45:235--245.

\bibitem[Jensen and Kowalski, 2012]{JensenKowalski2012}
Jensen, W.~A. and Kowalski, S.~M. (2012).
\newblock Response surfaces, blocking and split plots: An industrial experiment
  case study.
\newblock {\em Quality Engineering}, 24:531--542.

\bibitem[JMP, 2023]{JMP}
JMP (1989-2023).
\newblock {\em JMP$^\circledR$ Software. Version 10}.
\newblock SAS Institute Inc., Cary, North Carolina 27513, USA.

\bibitem[Jones and Goos, 2007]{JonesGoos2007}
Jones, B. and Goos, P. (2007).
\newblock A candidate-set-free algorithm for generating {D}-optimal split-plot
  designs.
\newblock {\em Applied Statistics}, 56:347--364.

\bibitem[Jones and Goos, 2009]{JonesGoos2009}
Jones, B. and Goos, P. (2009).
\newblock {D}-optimal design of split-split-plot experiments.
\newblock {\em Biometrika}, 96:67--82.

\bibitem[Jones and Goos, 2012]{JonesGoos2012}
Jones, B. and Goos, P. (2012).
\newblock {I}-optimal versus {D}-optimal split-plot response surface designs.
\newblock {\em Journal of Quality Technology}, 44:85--101.

\bibitem[Kowalski et~al., 2006]{Kowalskietal2006}
Kowalski, S.~M., Vining, G.~G., Montgomery, D.~C., and Borror, C.~M. (2006).
\newblock Modifying a central composite design to model the process mean and
  variance when there are hard-to-change factors.
\newblock {\em Journal of the Royal Statistical Society Series C: Applied
  Statistics}, 55:615--630.

\bibitem[Mead et~al., 2012]{Mead2012}
Mead, R., Gilmour, S.~G., and Mead, A. (2012).
\newblock {\em Statistical Principles for the Design of Experiments}.
\newblock Cambridge University Press.

\bibitem[Mylona et~al., 2020]{MylonaGilmouGoos2020}
Mylona, K., Gilmour, S.~G., and Goos, P. (2020).
\newblock Optimal blocked and split-plot designs ensuring precise pure-error
  estimation of the variance components.
\newblock {\em Technometrics}, 62:57--70.

\bibitem[Mylona et~al., 2014]{MylonaGoosJones2014}
Mylona, K., Goos, P., and Jones, B. (2014).
\newblock Optimal design of blocked and split-plot experiments for fixed
  effects and variance component estimation.
\newblock {\em Technometrics}, 56:132--144.

\bibitem[Sambo et~al., 2014]{Samboetal2014}
Sambo, F., Borroti, M., and Mylona, K. (2014).
\newblock A coordinate exchange two-phase local search algorithm for the {D}-
  and {I}-optimal designs of split-plot experiments.
\newblock {\em Computational Statistics \& Data Analysis}, 71:1193–1207.

\bibitem[Trinca and Gilmour, 2001]{TrincaGilmour2001}
Trinca, L. and Gilmour, S. (2001).
\newblock Multistratum response surface designs.
\newblock {\em Technometrics}, 43.

\bibitem[Trinca and Gilmour, 2015]{TrincaGilmour2015}
Trinca, L. and Gilmour, S. (2015).
\newblock Improved split-plot and multistratum designs.
\newblock {\em Technometrics}, 57.

\bibitem[Trinca and Gilmour, 2017]{TrincaGilmour2017}
Trinca, L. and Gilmour, S. (2017).
\newblock Split-plot and multi-stratum designs for statistical inference.
\newblock {\em Technometrics}, 59.

\end{thebibliography}

\end{document}